\documentclass[journal]{new-aiaa}
\usepackage[utf8]{inputenc}
\usepackage{float}
\usepackage{graphicx}
\usepackage{amsmath}
\DeclareMathOperator{\argmax}{arg\,max}
\usepackage[version=4]{mhchem}
\usepackage{siunitx}
\usepackage{longtable,tabularx}
\usepackage{subcaption}
\usepackage[dvipsnames]{xcolor}
\usepackage{comment}

\usepackage{xr}
\makeatletter
\newcommand*{\addFileDependency}[1]{% argument=file name and extension
  \typeout{(#1)}
  \@addtofilelist{#1}
  \IfFileExists{#1}{}{\typeout{No file #1.}}
}
\makeatother
\newcommand*{\myexternaldocument}[1]{
    \externaldocument{#1}
    \addFileDependency{#1.tex}
    \addFileDependency{#1.aux}
}
\myexternaldocument{SuppMat}
\title{Bayesian Estimation of Spectroscopic Parameters: \\ Application to the Atomic Nitrogen Bound-Bound System}

\author{Tae Woong Jeong\footnote{Ph.D. Student, Department of Aerospace Engineering.}}
\affil{Korea Advanced Institute of Science and Technology, Daejeon 34141, Republic of Korea}

\author{Sung Min Jo\footnote{Assistant Professor, Department of Mechanical and Aerospace Engineering, sungmin.jo@ucf.edu (Corresponding Author).}}

\affil{UCF Center of Excellence in Hypersonics and Space Propulsion (HYPERSPACE), \\ University of Central Florida, Orlando, FL 32816, USA}

\begin{document}

\maketitle

\begin{abstract}
Atomic nitrogen bound-bound radiation is a major component of the radiative heat flux on hypersonic vehicles entering nitrogen-dominated atmospheres, yet its prediction is limited by substantial parametric uncertainty in the published Einstein coefficients and Stark broadening coefficients. In the present study, these spectroscopic parameters are inferred and their uncertainty is quantified through Bayesian inversion of equilibrium spectral radiance measured in the NASA Ames Electric-Arc Shock Tube for two shots of the Test 62 campaign at shock speeds of 10.32 and 10.72 km/s. The inference is restricted to the post-shock equilibrium region, where the Boltzmann assumption closes the species population degree of freedom. The residual uncertainty in the post-shock temperature and species number densities is incorporated as a coupled nuisance parameter distribution. A hybrid principal component analysis and polynomial chaos expansion surrogate model and a likelihood formulated jointly over the two shots enable tractable Markov chain Monte Carlo sampling across multiple wavelength regions. Eighteen parameters in total, ten Einstein coefficients and eight Stark broadening coefficients, are inferred across eight wavelength regions, with posterior uncertainties significantly reduced relative to the prior literature bands. Forward propagation of the joint posterior through the stagnation-line flow field around a 3 m radius sphere at entry velocities of 10, 12, and 14 km/s demonstrates a reduction in the standard deviation of the predicted radiative heat flux by approximately a factor of five compared with the prior, in particular at 14 km/s, it drops from 10.4 to 1.94 W/cm$^{2}$.
\end{abstract}

\section{Introduction}
\lettrine{D}{uring} hypersonic atmospheric entry, a shock layer develops ahead of the vehicle, converting a substantial portion of the incoming kinetic energy into internal energy. This results in complex thermochemical nonequilibrium effects involving chemical reactions and energy exchanges. These processes expose the vehicle surface to intense convective and radiative heating. Therefore, the Thermal Protection System (TPS) should be designed based on an accurate prediction of these aerothermal loads to ensure mission success.

Furthermore, ionization reactions during hypersonic entry lead to the formation of a plasma region surrounding the vehicle, which can attenuate radio frequency (RF) signals and disrupt communication links \cite{kim2008analysis}. The electron number density that drives this attenuation arises from the same electronic state physics that governs radiative heat flux, so accurate modeling of these states is required for both phenomena.
In planetary atmospheres dominated by nitrogen, such as those of Earth and Titan, it is particularly important to precisely model the internal energy states of atomic and diatomic nitrogen, as these species are strong radiators and major sources of electron production.

Most studies on the electronic excitation of atomic nitrogen have been carried out at flow velocities greater than 10 km/s \cite{cruden2019analysis,tibere2018validation}. At such high speeds, ionization reactions generate a large number of electrons, and most electronic excitation processes are driven by electron impact collisions with atoms. As a result, when using a collisional-radiative (CR) model to predict the distribution of electronic states under these conditions, the contribution from heavy particle impact excitation can often be neglected or treated in a simplified manner.

In contrast, at flight speeds ranging from 5 to 8 km/s, ionization reactions are less dominant within the shock layer. Under these conditions, heavy particle impact excitation becomes more significant compared to cases with flight speeds above 10 km/s. However, it remains difficult to directly compute reliable reaction rates for heavy particle impact excitation using quantum chemical methods, owing to the residual limitations of the classical theories that describe such processes \cite{barklem2011inelastic}. Moreover, experimental measurements of such reactions are also difficult due to the relatively weak voltage signals they produce compared to those generated by electron impact excitation. 
A recent study developed a CR model for an $\mathrm{N_2}$--Ar mixture and compared the predicted population of $\mathrm{N}(3s\,^4P_{5/2})$ with shock tube measurements. The model was found to underestimate the population of $\mathrm{N}(3s\,^4P_{5/2})$ compared to the observed data \cite{aiken2023modeling}.
 
Bayesian inversion offers a principled framework for addressing these difficulties \cite{kennedy2001bayesian,swiler2006bayesian}, and the hypersonic community has demonstrated this technique for free-stream condition reconstruction \cite{cortesi2016rebuilding,cortesi2020forward}, and estimating parameters for gas-phase chemical-kinetics \cite{miki2012estimation,kuppa2025stochastic} and for Gas-Surface Interactions (GSI) \cite{upadhyay2011uncertainty, del2022surrogate, del2022stochastic, del2021bayesian, piro2025robust,rostkowski2019calibration,rostkowski2022effects,rostkowski2025non}. Before kinetic processes such as heavy particle impact excitation can be inferred this way, however, the radiance signal used as the inversion target, which is typically measured in the NASA Ames Electric-Arc Shock Tube (EAST) \cite{brandis2018shock, cruden2020measurement, brandis2016non, brandis2017titan}, must itself be modeled accurately. The radiance is governed by the Einstein coefficients and the line broadening parameters, both of which carry substantial parametric uncertainty in the published literature. The NIST Atomic Spectra Database assigns accuracy ratings of up to $\pm 50\%$ to individual Einstein coefficients for atomic nitrogen transitions \cite{wiese1996atomic}, and the Stark broadening coefficients reported by Johnston \emph{et al.} \cite{johnston2008spectrum} carry point-estimate uncertainty bands of up to $100\%$. These uncertainties propagate directly into any predicted equilibrium and nonequilibrium radiance and, by extension, into the radiative heat flux on a vehicle surface. This is independent of any uncertainty in the underlying kinetic rates. Quantifying and reducing the spectroscopic parameter uncertainty is therefore a prerequisite for any subsequent Bayesian inversion of kinetic processes, such as heavy particle and electron impact excitation.

Accordingly, the present study takes this prerequisite step. We quantify and reduce the parametric uncertainty in the Einstein coefficients and the Stark broadening coefficients for atomic nitrogen bound-bound transitions, leaving the analogous diatomic nitrogen problem to future work. Among the line broadening mechanisms, only the Stark broadening coefficients are inferred, because the Stark effect dominates in the high electron density regime relevant to the present conditions \cite{jo2019electronic}. The inference uses the equilibrium spectral radiance measured in EAST as the calibration data, restricted to the post-shock equilibrium region so that the Boltzmann assumption closes the species population degree of freedom and removes it from the set of inferred parameters. The residual uncertainty in the post-shock temperature and species number densities is incorporated as a nuisance parameter distribution that is itself coupled to the prior on the spectroscopic parameters. To exploit the available data robustly, the likelihood is constructed jointly over two EAST shots, T62-19 (10.32 km/s) and T62-21 (10.72 km/s), with a tempering weight that down-weights the data contribution when the per-shot maximum a posteriori estimates disagree. This provides the inference outcome that accounts for the consistency between the two-shot conditions.

To assess the impact of the inferred parametric uncertainty under practical hypersonic flight trajectories, the resulting posterior distributions are forward propagated through the stagnation line flow field around a 3 m radius sphere entering the Earth's atmosphere at velocities of 10, 12, and 14 km/s. Then, the resulting distribution of the radiative heat flux at the vehicle surface is compared with that obtained from the prior. The remainder of the paper is organized as follows. Section \ref{Sec:PhysicalModeling} describes the equilibrium radiance model, the inference setup, the surrogate construction, and the Bayesian framework. Section \ref{Sec:Results} presents the manufactured data construction, the surrogate validation, the posterior distributions of the inferred parameters, and the impact on radiative heating in the hypersonic stagnation line flows. Section \ref{Sec:Conclusions} summarizes the conclusions.

\section{Physical Modeling}\label{Sec:PhysicalModeling}

\subsection{Equilibrium Radiance of Atomic Nitrogen Bound-Bound Systems}

To model spectrally resolved radiance due to the atomic nitrogen bound-bound systems in thermochemical equilibrium, the emission coefficient $\epsilon_{\lambda}$ at a given wavelength $\lambda$ is defined as:

\begin{equation}
\epsilon_{\lambda}
=
\frac{h c}{4\pi}
\frac{N_k A_{ki}}{\lambda_{ki}}
\, \Phi_{\lambda},
\label{eq:eq1}
\end{equation}

\noindent
where $h$ and $c$ correspondingly denote the Planck constant and the speed of light. The subscripts $k$ and $i$ refer to the upper and lower electronic states, respectively. $A_{ki}$ is the Einstein coefficient for spontaneous emission at the transition wavelength $\lambda_{ki}$. $N_k$ is the upper electronic state population determined by a Boltzmann distribution: 

\begin{equation}
N_k
=
N_N \frac{g_k \exp\left(-E_k / \left(k_B T_{el}\right)\right)}{\sum_j g_j \exp\left(-E_j / \left(k_B T_{el}\right)\right)},
\label{eq:Nk}
\end{equation}

\noindent
where $N_N$ is the total number density of atomic nitrogen. $g_j$ and $E_j$ denote the electronic degeneracy and term energy of the $j$-th state, respectively. $k_B$ and $T_{el}$ are the Boltzmann constant and electronic temperature of atomic nitrogen.

In Eq. (\ref{eq:eq1}), $\Phi_{\lambda}$ represents the line broadening function based on a Voigt line profile proposed by Olivero and Longbothum \cite{OLIVERO1977233}. This includes Doppler broadening, classical natural broadening, pressure broadening, and Stark broadening. For the present condition of interest, the half-width at half-maximum (HWHM) of Stark broadening is orders of magnitude larger than those of the others \cite{potter2011modelling} due to free electrons, making it one of the major uncertain parameters in the present study. The HWHM of the Stark broadening $\Delta \lambda_S$ can be modeled as \cite{johnston2008spectrum}:

\begin{equation}
\Delta \lambda_S
=
\Delta \lambda_{S,0}
\left(
\frac{T_e}{10{,}000}
\right)^n
\left(
\frac{N_e}{1 \times 10^{16}\,\mathrm{cm^{-3}}}
\right)
\ \mathrm{nm},
\label{eq:eq2}
\end{equation}

\noindent
where $\Delta \lambda_{S,0}$ denotes the Stark broadening coefficient. $T_e$ and $N_e$ are the temperature and number density of free electrons, respectively. The exponent $n$ is typically taken as 1/3 for most nitrogen spectral lines \cite{park1982calculation}. It is important to note that the present study assumes a thermochemical equilibrium state, leading to $T_e = T_{el} = T$, where $T$ denotes the equilibrium temperature.

By neglecting the radiation cooling effect, the spectrally resolved equilibrium radiance $I_{\lambda}$ in a homogeneous medium that is artificially smeared due to the spectrometer can be calculated as:

\begin{equation}
I_{\lambda}=\int_{-\infty}^{+\infty} S_{\lambda} \left(1-e^{-\kappa_{\lambda} D}\right) \Phi_{\mathrm{ILS}}\left(\lambda-\lambda^{\prime}\right) d \lambda^{\prime},
\label{eq:radiance}
\end{equation}

\noindent
where $S_{\lambda}$ is the source function, which is the ratio of emission and absorption coefficients. $\kappa_{\lambda}$ and $D$ are the absorption coefficient and the shock tube diameter, respectively. In the present study, $\kappa_{\lambda}$ is determined by using $\epsilon_{\lambda}$ and $S_{\lambda}$ based on Kirchhoff's law. $\Phi_{\text{ILS}}$ is the instrumental line shape (ILS) function that is defined according to the specification of the spectrometer and the calibration ramp provided in the EAST measurement \cite{brandis2018shock}. It follows from Eqs.~(\ref{eq:eq1})--(\ref{eq:radiance}) that the Einstein coefficients $A_{ki}$ and the Stark broadening coefficients $\Delta \lambda_{S,0}$ are key uncertain parameters that directly influence the equilibrium radiance $I_{\lambda}$ of the atomic nitrogen bound-bound systems, therefore considered in the present Bayesian estimation.

\subsection{Overview of the Uncertainty Quantification Framework}
\begin{figure}[hbt!]
\centering
\includegraphics[width=0.8\textwidth]{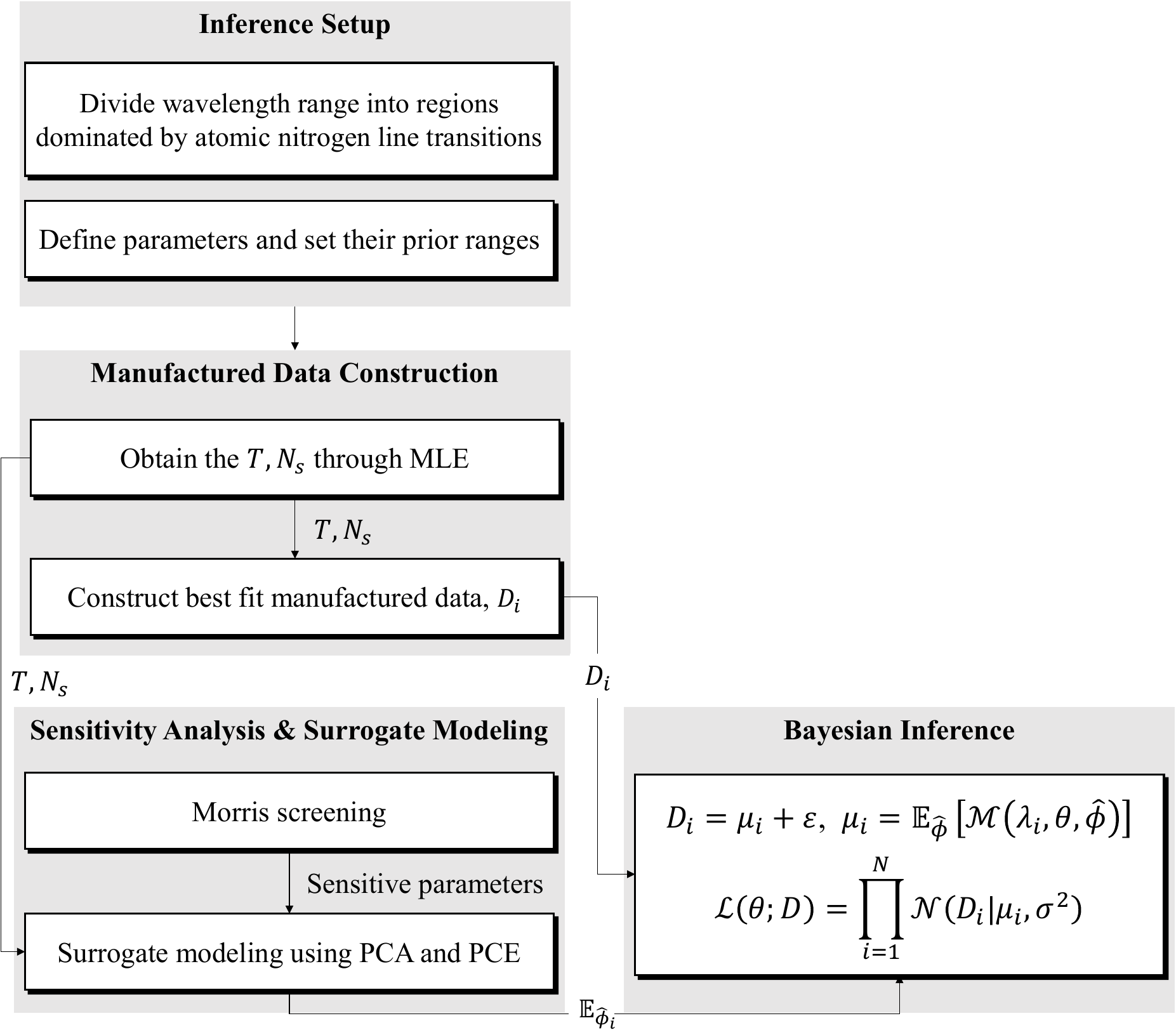}
\caption{Schematic of the present uncertainty quantification framework.}
\label{fig:fig1}
\end{figure}

To quantify the uncertainty of the spectroscopic parameters of the atomic nitrogen bound-bound system (\emph{i.e.}, $A_{ki}$ and $\Delta \lambda_{S,0}$), spectrally resolved equilibrium radiance spanning Vacuum-UV (VUV) to Infrared (IR) is used as the QoI. The in-house computational tools \textsc{plato} \cite{munafo2025plato,munafo2020computational} and \textsc{murp} \cite{jo2023multi} were employed to predict the QoI. The temperature ($T$) and number density of chemical species $s$ ($N_s$) in the equilibrium region behind a normal shock wave are predicted using \textsc{plato}. The spectrally resolved equilibrium radiance is then obtained from \textsc{murp} using the flow properties. In addition to generating the model output, various components required to solve the inversion problem, such as Bayesian inference, surrogate modeling, and sensitivity analysis, are implemented using the \textsc{UQpy}\cite{tsapetis2023uqpy} toolbox. Figure~\ref{fig:fig1} summarizes the overall uncertainty quantification framework adopted in the present study, which consists of four key steps. Further details for each step are provided in Sec. \ref{Inference Setup}--\ref{Bayesian Inference}.

\subsection{Inference Setup}\label{Inference Setup}
Since the parameters of interest are associated with individual atomic nitrogen lines, the wavelength domain is partitioned into regions where the features of the bound–bound transitions are dominant. Subsequently, the prior distribution ranges for each parameter are specified based on physical considerations.

\subsubsection{Measured Equilibrium Radiance Data}

As reference data for the physical model, measured equilibrium radiance data were taken from the Test 62 (T62) campaign of the EAST test series, which was conducted under pure nitrogen conditions \cite{brandis2018shock}. Among the test data, shots 19 (T62-19) and 21 (T62-21), with shock speeds of 10.32 km/s and 10.72 km/s, respectively, are selected, because those shots are regarded as benchmark-quality by the experimental group and have sufficiently high shock speeds to ensure a high signal-to-noise ratio in the measured equilibrium radiance of the atomic nitrogen bound-bound lines. Two shots are used so that a joint inference approach can be employed (see Sec. \ref{Bayesian Inference}) to mitigate the influence of shot-to-shot variability \cite{brandis2015uncertainty} and to infer spectroscopic parameters  consistent across different shock speeds.

\begin{figure}[hbt!]
\centering
\includegraphics[width=0.5\textwidth]{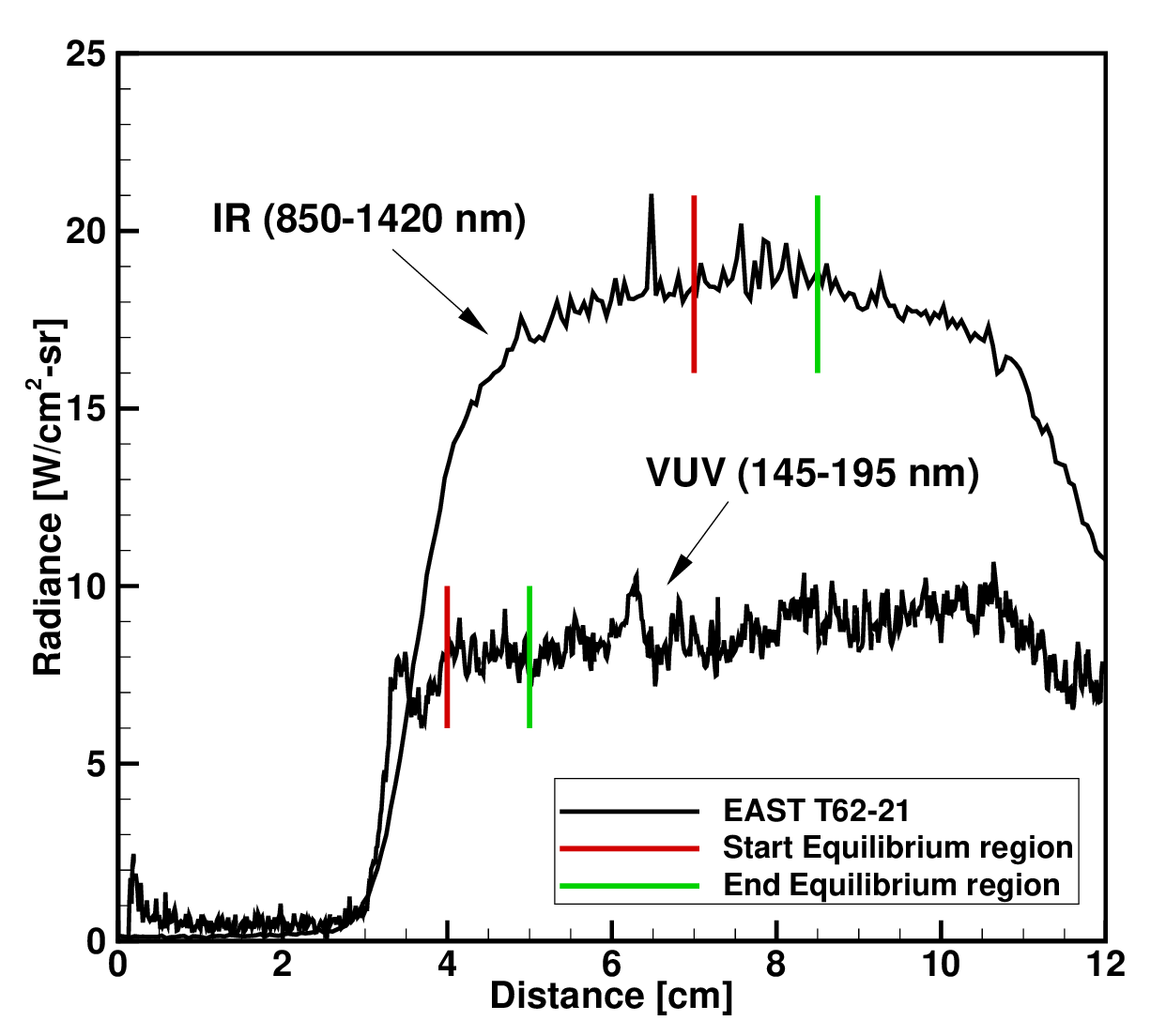}
\caption{Example of the equilibrium plateau of EAST T62-21 for VUV (level-3) and IR (level-4) ranges.}
\label{fig:fig2}
\end{figure}

Figure \ref{fig:fig2} shows the example of measured spatial radiance profiles of T62-21 in the VUV and IR ranges. The reference data for the Bayesian inference of the spectrally resolved equilibrium radiance are then obtained by taking a spatial average over the equilibrium region, which is marked by the two vertical lines.
In extracting the spectrally resolved equilibrium radiance from the experimental data, it is important to identify an appropriate equilibrium region along the spatial domain. In the previous work of Brandis \emph{et al.} \cite{brandis2015uncertainty} and Brandis \cite{brandis2010analysis}, identification of the equilibrium region of the post-shock radiance was classified into level-1 through level-5 according to how clearly the equilibrium plateau is formed, with level-5 indicating a longer and more constant plateau. In the present study, the considered spectral regions of the VUV (145--195 nm), Red (480--890 nm), and IR (850--1420 nm) are confirmed to have level-2 or higher classification for T62-19 and level-3 or higher for T62-21, resulting in the robust identification of the equilibrium radiance that can provide high-quality reference data for the present uncertainty quantification.

In the present study, the inference focuses on $A_{ki}$ and $\Delta \lambda_{S,0}$ associated with bound-bound transitions of atomic nitrogen. Accordingly, within the VUV, Red, and IR wavelength regions, the spectral domains where the features of the bound-bound transitions critically appear are further divided. The resultant partitioned wavelength regions considered in the present study for the Bayesian inference consist of 17 wavelength regions and are summarized in Table~\ref{tab:table1}.

\begin{table}[h!]
\centering
\caption{Partitioned wavelength regions considered for the present uncertainty quantification.}
\label{tab:table1}
\begin{tabular}{c c | c c}
\hline\hline
\textbf{Region} & \textbf{Wavelength (nm)} & \textbf{Region} & \textbf{Wavelength (nm)} \\
\hline
VUV-1 & 149--150    & IR-1 & 900--910 \\
VUV-2 & 173--175    & IR-2 & 915--923 \\
Red-1 & 640--652.5  & IR-3 & 933--955 \\
Red-2 & 659.5--667.5 & IR-4 & 975--1007.2 \\
Red-3 & 668--678    & IR-5 & 1007.5--1021 \\
Red-4 & 730--750    & IR-6 & 1045--1081 \\
Red-5 & 788--792    & IR-7 & 1117--1137 \\
Red-6 & 810--830    & IR-8 & 1196--1264 \\
Red-7 & 856--880    &      &           \\
\hline\hline
\end{tabular}
\end{table}

\subsubsection{Prior Distributions of Model Parameters\label{Sec:prior}}
In the present study, the Atomic Spectra Database (ASD) \cite{FIPS1402}, compiled by the National Institute of Standards and Technology (NIST), was used to obtain the baseline Einstein coefficients and their prior uncertainty ranges for the bound-bound radiative transitions of atomic nitrogen. The NIST ASD provides tabulated radiative transition probabilities for various atomic species. In 1996, Wiese \emph{et al.} \cite{wiese1996atomic} extensively revised and expanded the atomic radiative transition probabilities that had been published by the National Bureau of Standards (NBS) \cite{wiese1969atomic} about 30 years earlier.  
The update was carried out using both experimental measurements and calculated data, with the majority of revisions based on results from theoretical approaches. The latter was primarily obtained from the Opacity Project (OP) \cite{seaton1987atomic,
berrington1987atomic,yan1987atomic,fernley1987atomic,luo1989atomic,tully1990atomic} and the CIV3 code \cite{hibbert1993accurate,hibbert1991new,hibbert1991e1,bell1992oscillator,bell1994accurate,bell1995transition}, which are based on multiconfiguration calculations. In contrast to the CIV3 code, which computes transition probabilities for both individual lines and multiplets, the OP calculations only provide data for the multiplet. Therefore, for the OP results, the transition probabilities for individual lines were obtained using LS-coupling \cite{cowan2023theory,allen1973astrophysical} line strength fractions. Later, Wiese \emph{et al.} \cite{wiese2007improved} updated selected transition probabilities based on the results of multiconfiguration Hartree-Fock (MCHF) calculations performed by Tachiev and Fischer \cite{tachiev2002breit}.

Since these multiconfiguration approaches did not provide explicit estimates of parametric uncertainties in the calculated transition probabilities, the uncertainties had to be inferred from comparisons with available experimental data and other theoretical results.
Accordingly, Wiese \emph{et al.} \cite{wiese1996atomic} evaluated the uncertainties of the calculated Einstein coefficients for the transitions among low-lying energy states by comparing with the available experimental and theoretical data. This led to an accuracy level being assigned to each transition, ranging from AA to E, according to the criteria given in Table \ref{tab:table2}.
On the other hand, for transitions between high-lying energy states, only numerical calculations are available, with no supporting experimental measurements. In this case, Wiese \emph{et al.} \cite{wiese1996atomic}  assigned the accuracy levels based on the line strength at the multiplet level, and then determined the accuracy ratings for individual lines by stepwise downgrading from the rating of the corresponding multiplet using the same criteria as in Table \ref{tab:table2}. Based on this investigation, the prior uncertainty distributions of the Einstein coefficients considered in the present study are assigned following the criteria in Table \ref{tab:table2}.

\begin{table}[h!]
\centering
\caption{Accuracy levels and corresponding uncertainty ranges for the Einstein coefficients of atomic bound-bound radiative transition taken from the work of Wiese \emph{et al.} \cite{wiese1996atomic}.}
\label{tab:table2}
\begin{tabular}{c c}  
\hline\hline
\textbf{Accuracy level} & \textbf{Uncertainty range} \\ 
\hline
AA & uncertainties within $\pm 1\%$ \\
A  & uncertainties within $\pm 3\%$ \\
B  & uncertainties within $\pm 10\%$ \\
C  & uncertainties within $\pm 25\%$ \\
D  & uncertainties within $\pm 50\%$ \\
E  & uncertainties larger than $\pm 50\%$ \\
\hline\hline
\end{tabular}
\end{table}

Unlike the Einstein coefficients, which are defined for individual transitions, the Stark broadening coefficients are typically assigned at the multiplet level. Individual lines within the same multiplet share the same upper state electron configuration, characterized by the principal quantum number ($n$), orbital angular momentum quantum number ($l$), total spin quantum number ($S$), and total orbital angular momentum ($L$), and similarly share the same lower state electron configuration ($n'$, $l'$, $S'$, $L'$).
They differ only in the total angular momentum quantum number ($J$). Consequently, the individual lines comprising a single multiplet have the same sum of cross sections for inelastic scattering. Since the Stark broadening width of a line due to electrons is calculated using the sum of the cross sections for the inelastic scattering, the widths of the individual lines within a multiplet can be treated as identical \cite{wiese1982regularities}.

\begin{figure}[hbt!]
\centering
\includegraphics[width=0.5\textwidth]{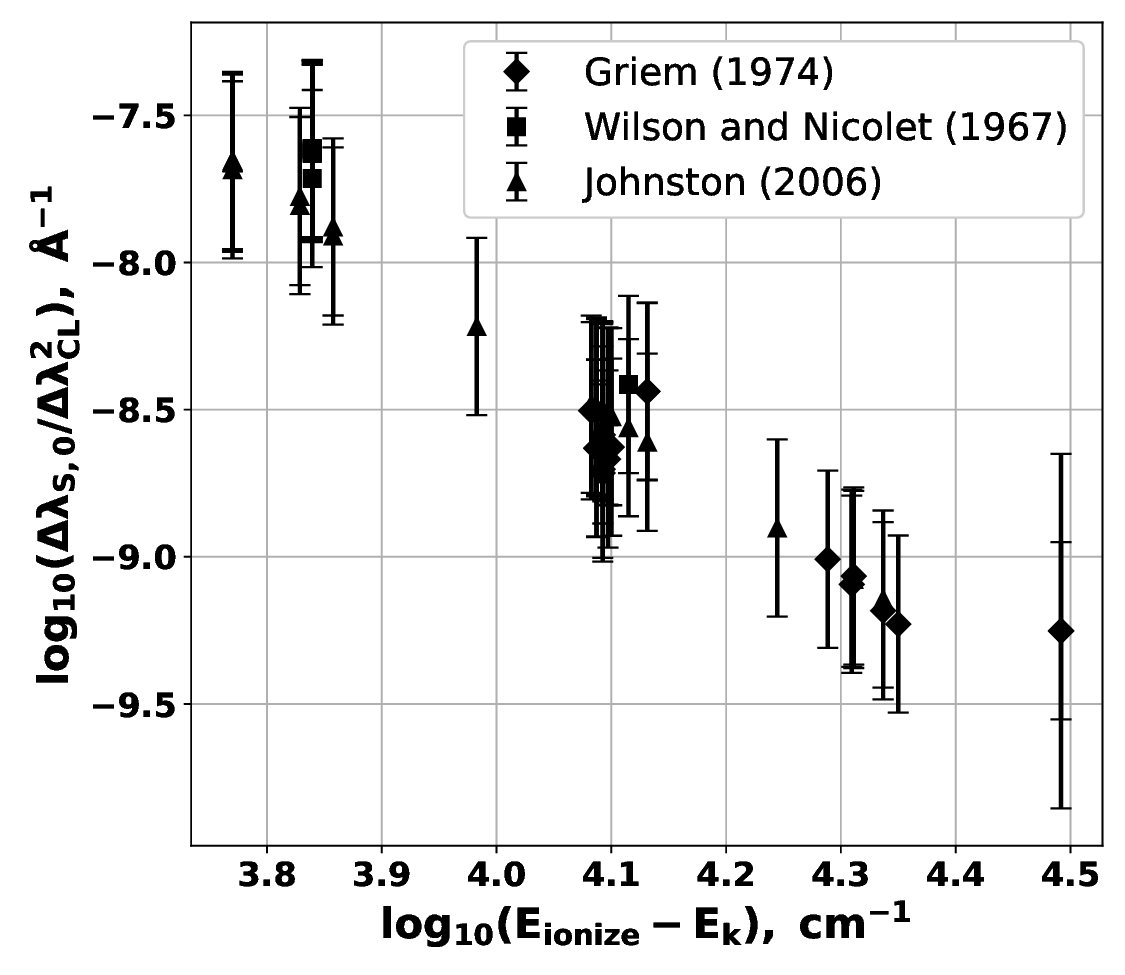}
\caption{Prior uncertainties of Stark broadening coefficients.}
\label{fig:fig3}
\end{figure}

In this vein, the present Bayesian inference for the Stark broadening coefficient $\Delta \lambda_{S,0}$ is performed at the multiplet level using the nominal values defined accordingly. The data taken from the previous studies by Griem \cite{1974slbp.book.....G}, Wilson and Nicolet \cite{wilson1967spectral}, and Johnston \cite{johnston2006nonequilibrium} are considered as the baseline. The first two sources were prioritized over the fitting function proposed by Johnston \cite{johnston2006nonequilibrium}. The prior uncertainty ranges of $\Delta \lambda_{S,0}$ were determined based on the values reported in the previous study by Johnston \emph{et al.} \cite{johnston2008spectrum}. Specifically, the maximum reported uncertainty of 100\% was adopted for all coefficients, except for those associated with the VUV-2 region. For the latter case, a larger prior uncertainty of 300\% was assigned based on a preliminary analysis using the baseline Einstein coefficient and $\Delta \lambda_{S,0}$, which indicated substantially higher parametric uncertainty in this spectral range. Similar to the Einstein coefficients, a uniform prior distribution is assigned to $\Delta \lambda_{S,0}$ using the corresponding nominal value and uncertainty range. Figure~\ref{fig:fig3} presents the prior uncertainties of $\Delta \lambda_{S,0}$, where the nominal values are indicated by black markers. In the figure, $\lambda_{\mathrm{CL}}$ is the line-center wavelength, while $E_{\mathrm{ionize}}$ and $E_k$ denote the ionization potential of atomic nitrogen and the upper state term energy, respectively.

\subsection{Manufactured Data Construction}\label{Manufactured Data Construction}

In the EAST measured data, individual atomic nitrogen bound-bound lines are smeared and merged together due to the effect of the ILS function associated with the spectrometer as described in Eq. (\ref{eq:radiance}). This poses a challenge for leveraging the EAST measured data within the present uncertainty quantification framework, which requires that the individual transition lines are recognizable. To overcome this limitation, manufactured data are numerically constructed in the present study and used to replace the actual experimental measurements, enabling the stochastic inference for the individual lines.

To construct the manufactured data, Maximum Likelihood Estimation (MLE) is used to determine a set of spectroscopic parameters that best reproduces the EAST measurement within their prior ranges. The QoI used for Bayesian inference is the spectral radiance, which strongly depends on temperature $T$ and species number densities \cite{brandis2012validation}, in this study, $N_N$ and $N_e$. The post-normal shock equilibrium flow state can be calculated by combining the Rankine-Hugoniot shock jump condition with the Gibbs free energy minimization \cite{anderson1989hypersonic}. However, due to the characteristics of the shock tube environment, such as shock deceleration due to boundary layer growth and limited test time, it remains uncertain that the flow inside the EAST facility reaches an ideal thermochemical equilibrium state. Furthermore, the previous studies by Brandis \emph{et al.} \cite{brandis2015uncertainty} and Cruden \cite{cruden2012electron} compared the measured and calculated electron number density profiles to assess the degree of equilibrium in the EAST facility, and found that the agreement deteriorates as the shock speed decreases. In this vein, to generate the manufactured data, the temperature and species number densities are treated as nuisance parameters: they are propagated as uncertain quantities so that their effect is incorporated into the Bayesian inference of the spectroscopic parameters of interest, but they are not themselves the targets of the calibration.

The uncertainty in the nuisance parameters (\emph{i.e.}, $T$, $N_N$, and $N_e$) is coupled to that of the target inference parameters, the Einstein and Stark broadening coefficients. To incorporate this coupling effect, an MLE is performed for an individual sampled Einstein and Stark broadening coefficients from their prior distributions, defined in Sec. \ref{Sec:prior}, to best estimate the corresponding nuisance parameters that reproduce the EAST measurement. This leads to joint distributions of $T$, $N_N$, and $N_e$, as presented in Sec. \ref{Sec:DataConstruction}, that incorporate the uncertainty of nuisance parameters affected by the target inference parameters.
By defining the inference parameters (\emph{i.e.}, the Einstein and Stark broadening coefficients) as $\theta$ and the nuisance parameters as $\phi$, the MLE expression can be formulated as follows:

\begin{equation}
\begin{aligned}
\theta^{k} &\sim \mathcal{P}(\theta), \qquad k = 1, \ldots, n_{\theta}, \\
\hat{\phi}^{k} &= \argmax_{\phi}\!\mathcal{P}(D \mid \theta^{k}, \phi),
\end{aligned}
\label{eq:MLE}
\end{equation}

\noindent
where $n_{\theta}$ is the number of samples drawn from the prior. $\mathcal{P}(\theta)$ denotes the prior probability density functions (PDFs) of $\theta$, sampled using Latin Hypercube Sampling (LHS) \cite{mckay2000comparison}. $\hat{\phi}^{k}$ represents the nuisance parameter inferred by MLE for a given $\theta^{k}$, thereby yielding the joint distributions of $T$, $N_N$, and $N_e$.
The reference data $D$ corresponds to the spectrally integrated intensity calculated using the EAST measurements in each partitioned wavelength range, defined in Table~\ref{tab:table1}. It is worth noting that $\hat{\phi}^{k}$ is also used for constructing surrogate models, described in Sec. \ref{Sec:SA_SM}. As the final step in constructing the manufactured data, $\hat{\phi}^{k}$ is propagated through the physical model by evaluating Eq.~(\ref{eq:radiance}) with the ILS function applied. Then, for each wavelength range, the ILS-convolved equilibrium spectral radiance that minimizes the sum of squared residuals (SSR) relative to the EAST data is identified. The corresponding equilibrium spectral radiance before applying the ILS convolution is then used as the manufactured data of the uncertainty quantification framework. This allows the elimination of the ILS function from the Bayesian inference step, thereby enabling consideration of individual line transitions without the apparatus-driven smearing.

\subsection{Sensitivity Analysis and Surrogate Modeling \label{Sec:SA_SM}}

Among the parameters within each partitioned wavelength region, those most sensitive to the integrated intensity are selected through sensitivity analysis. Subsequently, a surrogate model is constructed for the individual wavelength range by using a combination of Principal Component Analysis (PCA) and Polynomial Chaos Expansion (PCE). The output of the surrogate model is the spectral radiance obtained by taking the expectation on the nuisance parameter distributions of the temperature and the number density that are constructed through MLE as described in Sec. \ref{Manufactured Data Construction}.

In the partitioned wavelength ranges considered in this study, a total of 266 Einstein coefficients and 64 Stark broadening coefficients are involved. Because it is impractical to calibrate all of these parameters using Bayesian inference with the limited dataset, sensitivity analysis is performed separately for each wavelength range to identify the most influential parameters. A Morris screening method \cite{saltelli2008global} is employed for the sensitivity analysis, with the spectrally-integrated intensity selected as the QoI. For each partitioned wavelength region, the QoI is then defined as
\begin{equation}
Y(\boldsymbol{\theta}) =
\int_{\lambda_a}^{\lambda_b} I_\lambda(\boldsymbol{\theta}) \, d\lambda,
\end{equation}where $\lambda_a$ and $\lambda_b$ denote the lower and upper bounds of the partitioned wavelength region, respectively. The elementary effect of the $k$-th parameter along the $j$-th Morris trajectory can be computed as
\begin{equation}
EE_k^{(j)} =
\frac{
Y(\theta_1^{(j)},\ldots,\theta_k^{(j)}+\Delta,\ldots,\theta_d^{(j)})
-
Y(\theta_1^{(j)},\ldots,\theta_k^{(j)},\ldots,\theta_d^{(j)})
}{\Delta},
\end{equation}
where $\Delta$ is the prescribed perturbation size in the normalized parameter space and $d$ is the number of uncertain parameters in the corresponding wavelength region.
For each trajectory, the elementary effects are computed, and their mean ($\mu_k$), absolute mean ($\mu_k^*$), and standard deviation ($\sigma_k$) are evaluated as
\begin{equation}
\mu_k =
\frac{1}{r}
\sum_{j=1}^{r}
EE_k^{(j)},
\qquad
\mu_k^* =
\frac{1}{r}
\sum_{j=1}^{r}
\left| EE_k^{(j)} \right|,
\end{equation}
and
\begin{equation}
\sigma_k =
\sqrt{
\frac{1}{r}
\sum_{j=1}^{r}
\left(
EE_k^{(j)}-\mu_k
\right)^2
}.
\end{equation}
Here, $r$ denotes the number of Morris trajectories. A parameter with a large $\mu_k^*$ indicates a strong linear and additive influence on the QoI, whereas a large $\sigma_k$ indicates that the parameter exhibits nonlinear or interactive effects.
Accordingly, if a given parameter exhibits a large $\mu_k^*$ or $\sigma_k$ value within a specific wavelength region, it is identified as a sensitive parameter for that region. For each wavelength region, once the sensitive parameters are identified, only those parameters are used to construct the surrogate model and to perform the subsequent Bayesian inference.

Obtaining the posterior distribution via Bayesian inference can be computationally expensive. Furthermore, as the number of parameters increases, the parameter space the MCMC chain must explore becomes larger, which can slow down convergence. To overcome this, a surrogate model has been constructed in the present study to enable fast prediction of the QoI during the Bayesian inference.
To construct the surrogate model, PCA was first applied to the spectrally resolved radiance space defined on the wavelength grid to reduce the dimensionality of the model output space. Assuming that there are $N$ samples and $M$ wavelength points, the data matrix $\mathbf{Z}$ can be organized as follows:

\begin{equation}
\mathbf{Z} = 
\begin{bmatrix}
\mathbf{f}^{(1)} \\
\vdots \\
\mathbf{f}^{(N)}
\end{bmatrix}
\in \mathbb{R}^{N \times M}, \quad \text{where } \mathbf{f} \in \mathbb{R}^M, \, \mathbf{f} = \mathbb{E}_{\hat{\phi}}\!\left[ \mathcal{M}(\lambda_M, \theta, \hat{\phi}) \right].
\end{equation}

\noindent
The elements of the data matrix $\mathbf{Z}$ are given by 
$\mathbb{E}_{\hat{\phi}}\!\left[ \mathcal{M}(\lambda_M, \theta, \hat{\phi}) \right]$, 
where $\mathbb{E}$ denotes the expectation, and $\mathcal{M}$ is the model. This means that, for each $\theta$, the data matrix is formed by taking the expectation over the distributions of the nuisance parameters obtained in the manufactured data construction step. Through this process, the uncertainties of the nuisance parameters can be incorporated in the inference of the Einstein coefficients and Stark broadening coefficients.

Next, let $\mathbf{Z}_0$ denote the sample-wise mean row vector. The mean-subtracted data matrix $\overline{\mathbf{Z}} = \mathbf{Z} - \mathbf{Z}_0$ is then used to compute the covariance matrix. By performing eigen decomposition on this covariance matrix, the eigenvectors and their corresponding eigenvalues are obtained. These eigenvectors are assembled into a matrix $\mathbf{V}$, with its columns arranged in descending order of their eigenvalues. Since eigenvectors associated with larger eigenvalues indicate directions of significant variance, selecting a subset of leading eigenvectors can capture the dominant modes of variation in the data. Through this PCA, the centered data defined over the wavelength grid is projected onto a lower-dimensional subspace spanned by the leading eigenvectors, thereby reducing the dimensionality.

Then, the reduced data $\mathbf{Y} = \overline{\mathbf{Z}}\mathbf{V}$, combined with the corresponding parameters, serves as the training set for the PCE surrogate model. In this process, the reduced output matrix $\mathbf{Y}$ is represented using polynomial basis functions:
\begin{equation}
\mathbf{Y}_{\mathrm{PCE}} = \sum_{\alpha \in \mathrm{basis}} c_\alpha \Psi_\alpha(\mathbf{X}),
\end{equation}
where $c_\alpha$ denotes the expansion coefficients and $\Psi_\alpha$ represents the polynomial basis functions of the input parameters $\mathbf{X}$. The lower-dimensional output predicted by the PCE surrogate can be reconstructed to the original wavelength-grid dimension by multiplying by the transpose of the eigenvector matrix and re-adding the sample mean:
\begin{equation}
\mathbf{Z}_{\mathrm{recon}} = \mathbf{Y}_{\mathrm{PCE}}\mathbf{V}^{T} + \mathbf{Z}_0.
\end{equation}
In this study, Legendre polynomials \cite{dattoli1999generalized} were employed as the basis functions because the input parameters have uniform prior distributions. For all cases, the maximum polynomial degree was set to three. As a result, $\mathbf{Z}_{\mathrm{recon}}$ is employed to perform Bayesian inference.

\subsection{Bayesian Inference}\label{Bayesian Inference}

In the present study, the spectroscopic parameters of interest are estimated through Bayesian inference. The posterior distribution of the uncertain parameters is obtained via Bayes' theorem as:
\begin{equation}
\label{eq:eq3}
\mathcal{P}(\theta \mid D) = \frac{\mathcal{L}(\theta; D)^{w} \, \mathcal{P}(\theta)}{\mathcal{P}(D)},
\end{equation}

\noindent
where $\mathcal{P}(\theta \mid D)$ and $\mathcal{P}(\theta)$ denote the posterior and prior PDFs, respectively. 
$D$ represents the spectrally resolved intensity from the manufactured data. $\mathcal{P}(D)$ acts as the evidence term serving as a normalization constant that ensures the posterior distribution integrates to one. For the present Bayesian inference, a Markov Chain Monte Carlo (MCMC) algorithm \cite{metropolis1953equation} is used to obtain the posterior PDF of the parameters by drawing samples using a Delayed Rejection Adaptive Metropolis (DRAM) method \cite{haario2006dram}. The core idea behind MCMC is to construct a Markov chain that asymptotically converges to the target distribution, $\mathcal{P}(\theta \mid D)$, thereby generating samples that follow this posterior distribution over time.

The Bayesian inference procedure described in this section is independently applied for each partitioned wavelength region in the same manner. In Eq. (\ref{eq:eq3}), \(\mathcal{L}(\theta; D)^w\) is the likelihood function with a tempering parameter $w$ specified using an additive error model defined as:

\begin{equation}
D_i = \mathbb{E}_{\hat{\phi}}\!\left[ \mathcal{M}(\lambda_i, \theta, \hat{\phi}) \right]
 + \varepsilon, \quad \varepsilon \sim \mathcal{N}(0, \sigma^2),
 \label{eqn:additive_model}
\end{equation}

\noindent
where the error $\varepsilon$ is assumed to follow a zero-mean Gaussian distribution with a variance $\sigma^2$, and the errors at different wavelength grid points $i$ are assumed to be statistically independent of one another. The mismatch between the model output and the reference data is attributed to both model inadequacy and data noise \cite{higdon2004combining}. By treating these two error sources as independent, the total variance can then be written as $\sigma^{2} = \sigma^{2}_{m} \text{ (\emph{i.e.}, model inadequacy)} + \sigma^{2}_{n} \text{ (\emph{i.e.}, data noise)}.$ We postulate that $\sigma^{2}_{n}$ represents the variance induced during the construction of the manufactured data. Due to the uncertainty distributions of temperature and number density (\emph{i.e.}, nuisance parameters) within a given wavelength region, variance in the spectral radiance is induced when constructing the manufactured data. The quantity $\sigma^{2}_{n}$ is then obtained by averaging the per-grid-point variance over the given wavelength range.
Then \(\mathcal{L}(\theta; D)^w\) is expressed as:

\begin{equation}
\mathcal{L}(\boldsymbol{\theta};\mathbf{D}_{19},\mathbf{D}_{21})^w
=
\left[
\prod_{i=1}^{P}
\frac{1}
{\sqrt{(2\pi)^2 \det(\boldsymbol{\Sigma})}}
\exp
\left(
-\frac{1}{2}
\left(
\mathbf{D}_i-\mathbf{E}(\lambda_i,\boldsymbol{\theta})\right)^T\boldsymbol{\Sigma}^{-1}\left(\mathbf{D}_i-\mathbf{E}(\lambda_i,\boldsymbol{\theta})\right)\right)\right]^w ,
\end{equation}

\begin{comment}
\begin{equation}
\mathcal{L}_r(\theta_r;D_{19,r},D_{21,r})^{w_r}
=
\left[
\prod_{i=1}^{P_r}
\frac{1}{\sqrt{(2\pi)^2 \det(\Sigma_r)}}
\exp\left(
-\frac{1}{2}
\left(\mathbf{D}_{r,i}-\mathbf{E}_r(\lambda_{r,i},\theta_r)\right)^T
\Sigma_r^{-1}
\left(\mathbf{D}_{r,i}-\mathbf{E}_r(\lambda_{r,i},\theta_r)\right)
\right)
\right]^{w_r},
\end{equation}
\end{comment}

\begin{equation}
\mathbf{D}_i - \mathbf{E}(\lambda_i,\theta)
=
\begin{pmatrix}
D_{i,19} - \mathbb{E}_{\hat{\phi}_{19}}\!\left[\mathcal{M}_{19}(\lambda_i,\theta,\hat{\phi}_{19})\right] \\
D_{i,21} - \mathbb{E}_{\hat{\phi}_{21}}\!\left[\mathcal{M}_{21}(\lambda_i,\theta,\hat{\phi}_{21})\right]
\end{pmatrix},
\qquad
\boldsymbol\Sigma
=
\begin{pmatrix}
\sigma_{19}^{2} & 0 \\
0 & \sigma_{21}^{2}
\end{pmatrix},
\end{equation}

\noindent
where $P$ is the number of grid points in the partitioned wavelength range, and the subscripts $19$ and $21$ correspond to the EAST shot~19 and shot~21 conditions, respectively. The variances $\sigma^2_{m,19}$ and $\sigma^2_{m,21}$ in the likelihood function are treated as unknown quantities that are inferred simultaneously along with $\theta$. The discrepancy term included in the likelihood function is computed using the manufactured data and the model output obtained in the previous steps. 

The tempering parameter $w$ was introduced to control the dataset's influence during the joint inference. When $w = 1$, the formulation reduces to the standard Bayesian update, whereas for $0 < w < 1$, the likelihood is tempered, resulting in a reduced influence of the data \cite{ibrahim2000power, bissiri2016general}. 
In the present study, the value of $w$ is determined based on the discrepancy between the MAP (Maximum A Posteriori) estimates obtained from the EAST shot 19 and shot 21 conditions.
Specifically, the difference between the MAP estimates from the two shot conditions is normalized by the prior range of each parameter as:
\begin{equation}
w_k = 1 - 
\frac{
\left|\hat{\theta}_{19,k} - \hat{\theta}_{21,k}\right|
}{
\theta_{k,\max} - \theta_{k,\min}
},
\end{equation}
\noindent
where $k=1,\ldots,d$ denotes the index of the inferred parameter, and $d$ is the number of inferred parameters in the corresponding wavelength region. 
Here, $\hat{\theta}_{19,k}$ and $\hat{\theta}_{21,k}$ are the MAP estimates of the $k$-th parameter obtained using only one of the EAST measurements either shot 19 or shot 21, respectively, while $\theta_{k,\min}$ and $\theta_{k,\max}$ denote the lower and upper bounds of the prior distribution of the $k$-th parameter. 
The value of $w_k$ is thus constrained to the range $[0,1]$. The representative tempering parameter applied to the likelihood for a given wavelength range is then defined as the minimum consistency weight across the inferred parameters:
\begin{equation}
w = \min_{k=1,\ldots,d} \left(w_k\right).
\end{equation}
This allows a conservative choice reducing the influence of the data when at least one inferred parameter exhibits inconsistency between the two shot conditions.
This formulation is motivated by the idea that the consistency between independent data sources reflects their reliability in constraining the parameter posterior distributions. As a result, when the two-shot conditions provide consistent information, the influence of the data is strengthened, whereas inconsistent information leads to a reduced contribution of the data in the inference process.

It is important to note that the Einstein coefficients and Stark broadening coefficients differ by roughly 8--10 orders of magnitude.
With parameters that differ substantially in scale, the covariance matrix of the proposal distribution used in the MCMC algorithm may fail to remain positive definite, thereby impeding the evolution of the MCMC chains. This is because when the covariance matrix is not positive definite, Cholesky decomposition cannot be performed, and consequently, the multivariate normal density cannot be evaluated. To avoid such a situation, all parameters are inferred in the logarithmic space in the present study. This approach enables the chains to converge within a relatively small number of iterations for all wavelength regions.

\subsection{Calculation Conditions and Assumptions}

Table~\ref{tab:table1} presents a total of 17 partitioned wavelength regions. Ideally, inference would be performed on all parameters within each of these regions, but this is not feasible, primarily because the inversion problem cannot be solved for wavelength regions in which the measured radiance does not lie within the model-output variance band induced by the prior distributions of the parameters. Accordingly, for each partitioned wavelength region, the associated $\theta$ are sampled from their prior distributions and propagated through the model. The inference is then conducted only for wavelength regions in which the EAST measurement profiles lie within the variance band of the model output produced by the priors. In contrast, the regions that fail this check are excluded from the present analysis.
More wavelength regions were excluded from the inference in the IR range than in the VUV and Red ranges. This may be due to the lower spectrometer resolution in the IR region, which makes the measurements more susceptible to uncertainties from the background continuum \cite{brandis2010analysis, brandis2015uncertainty} and data noise. As a result, 18 parameters are inferred in total across 8 different wavelength ranges, as summarized in Table \ref{tab:table3}.

\begin{table}[h]
\centering
\caption{Summary of inferred spectroscopic parameters for the selected wavelength regions.}
\label{tab:table3}
\renewcommand{\arraystretch}{0.9}

\begin{tabular}{c c c c c c}
\hline\hline
\textbf{Region} & \textbf{Symbol} & \textbf{Transition / $\lambda_{CL}$} 
& \textbf{Nominal value} & \textbf{Accuracy level} & \textbf{Prior (log-space)} \\
\hline
% ================== VUV_2 ==================
\textbf{VUV-2} & $\Delta\lambda_{S,0}^{(2)}$ & $N(5)\rightarrow N(3)$ 
& $1.70\times10^{-3}\,\mathring{A}$ & $\pm 300\%$ \cite{johnston2008spectrum}& $U[-7.76,\,-4.99]$ \\
\hline
% ================== Red_1 ==================
\textbf{Red-1} & $A_{ki}^{(308)}$ & $648.449\text{ nm}$ 
& $4.90\times10^{6}\,\mathrm{s^{-1}}$ & C \cite{tachiev2002breit}& $U[15.18,\,15.63]$ \\
               & $A_{ki}^{(309)}$ & $648.5545\text{ nm}$ 
& $3.67\times10^{6}\,\mathrm{s^{-1}}$ & C \cite{tachiev2002breit}& $U[14.89,\,15.34]$ \\
               & $A_{ki}^{(310)}$ & $648.66\text{ nm}$ 
& $4.20\times10^{6}\,\mathrm{s^{-1}}$ & C \cite{tachiev2002breit}& $U[15.03,\,15.47]$ \\
               & $\Delta\lambda_{S,0}^{(4)}$ & $N(29)\rightarrow N(8)$ 
& $9.90\times10^{-1}\,\mathring{A}$ & $\pm 100\%$ \cite{johnston2008spectrum} & $U[-0.70,\,0.68]$ \\
\hline
% ================== Red_2 ==================
\textbf{Red-2} & $A_{ki}^{(323)}$ & $664.6797\text{ nm}$ 
& $3.49\times10^{6}\,\mathrm{s^{-1}}$ & D \cite{tachiev2002breit}& $U[14.66,\,15.47]$ \\
               & $A_{ki}^{(327)}$ & $665.5295\text{ nm}$ 
& $2.74\times10^{6}\,\mathrm{s^{-1}}$ & D \cite{tachiev2002breit}& $U[14.42,\,15.23]$ \\
\hline
% ================== Red_4 ==================
\textbf{Red-4} & $A_{ki}^{(372)}$ & $742.5686\text{ nm}$ 
& $5.64\times10^{6}\,\mathrm{s^{-1}}$ & B \cite{tachiev2002breit}& $U[15.45,\,15.64]$ \\
               & $A_{ki}^{(376)}$ & $744.4348\text{ nm}$ 
& $1.19\times10^{7}\,\mathrm{s^{-1}}$ & B \cite{tachiev2002breit}& $U[16.20,\,16.39]$ \\
               & $A_{ki}^{(380)}$ & $747.0369\text{ nm}$ 
& $1.96\times10^{7}\,\mathrm{s^{-1}}$ & B \cite{tachiev2002breit}& $U[16.70,\,16.89]$ \\
               & $\Delta\lambda_{S,0}^{(15)}$ & $N(10)\rightarrow N(4)$ 
& $4.77\times10^{-2}\,\mathring{A}$ & $\pm 100\%$ \cite{johnston2008spectrum} & $U[-3.74,\,-2.35]$ \\
\hline
% ================== Red_6 ==================
\textbf{Red-6} & $\Delta\lambda_{S,0}^{(21)}$ & $N(9)\rightarrow N(4)$ 
& $4.42\times10^{-2}\,\mathring{A}$ & $\pm 100\%$ \cite{johnston2008spectrum} & $U[-3.81,\,-2.43]$ \\
\hline
% ================== Red_7 ==================
\textbf{Red-7} & $\Delta\lambda_{S,0}^{(22)}$ & $N(12)\rightarrow N(5)$ 
& $7.28\times10^{-2}\,\mathring{A}$ & $\pm 100\%$ \cite{johnston2008spectrum} & $U[-3.31,\,-1.93]$ \\
               & $\Delta\lambda_{S,0}^{(25)}$ & $N(8)\rightarrow N(4)$ 
& $4.48\times10^{-2}\,\mathring{A}$ & $\pm 100\%$ \cite{johnston2008spectrum} & $U[-3.80,\,-2.41]$ \\
\hline
% ================== IR_5 ==================
\textbf{IR-5} & $\Delta\lambda_{S,0}^{(40)}$ & $N(17)\rightarrow N(8)$ 
& $2.21\times10^{-1}\,\mathring{A}$ & $\pm 100\%$ \cite{johnston2008spectrum} & $U[-2.20,\,-0.82]$ \\
\hline
% ================== IR_8 ==================
\textbf{IR-8} & $A_{ki}^{(633)}$ & $1246.4662\text{ nm}$ 
& $1.82\times10^{7}\,\mathrm{s^{-1}}$ & B \cite{tachiev2002breit}& $U[16.62,\,16.81]$ \\
               & $A_{ki}^{(634)}$ & $1247.3027\text{ nm}$ 
& $2.18\times10^{7}\,\mathrm{s^{-1}}$ & B \cite{tachiev2002breit}& $U[16.80,\,16.99]$ \\
               & $\Delta\lambda_{S,0}^{(61)}$ & $N(18)\rightarrow N(11)$ 
& $4.06\times10^{-1}\,\mathring{A}$ & $\pm 100\%$ \cite{johnston2008spectrum} & $U[-1.59,\,-0.21]$ \\
\hline\hline
\end{tabular}
\end{table}

Table~\ref{tab:table3} lists the parameters selected for the inference through the sensitivity analysis in each wavelength region. The numbers in parentheses shown as superscripts for each symbol represent the indices corresponding to the atomic nitrogen line transitions in the computational model database. The line transitions in the database were grouped at the multiplet level for $\Delta \lambda_{S,0}$, resulting in a total of 64 multiplets. Thus, $\Delta \lambda_{S,0}^{(40)}$ denotes the Stark broadening coefficient for 40th multiplet. The third column of the table lists the multiplet lines for $\Delta \lambda_{S,0}$ and the line-center wavelengths for $A_{ki}$. There are 261 electronic energy levels for atomic nitrogen obtained from the NIST ASD, and these can be further grouped into 38 electronic levels based on the principal quantum numbers, $n_p$ \cite{jo2019electronic}. Therefore, if the third column indicates $\mathrm{N}(5) \rightarrow \mathrm{N}(3)$, it can be interpreted as a transition from the 5th grouped level to the 3rd. For brevity, the complete level information is provided in Table S1 in the Supporting Material. The values in the fourth and fifth columns were determined according to the description provided in Sec. \ref{Sec:prior}. The prior ranges for all parameters are defined in log-space, using the natural logarithm, as shown in the last column.

Before proceeding to the results, it is necessary to clarify additional modeling assumptions regarding the coupling between radiation and the flow field. In the experimentally measured spatial radiance shown in Fig. \ref{fig:fig2}, a gradual decay is observed following the plateau. This behavior may arise from the arrival of the contact surface and from radiative cooling. The latter is a non-adiabatic effect that can be captured through coupling between the flow and the radiative transfer equation \cite{cruden2012electron,cruden2014absolute}. In addition, boundary layer growth along the shock tube wall can cause absorption of part of the emitted light, thereby affecting the experimental results. However, the region where radiative cooling occurs lies outside the area of interest considered in this study, and absorption in the boundary layer is significant only below 140 nm \cite{brandis2012validation}. Therefore, when calculating the spectral radiance with the present model, the effects of flow-radiation coupling and boundary layer growth are neglected.

\section{Results and Discussion}\label{Sec:Results}
\subsection{Manufactured Data Construction\label{Sec:DataConstruction}}

\begin{figure}[h]
\centering
\begin{subfigure}{0.48\textwidth}
    \centering
    \includegraphics[width=\textwidth]{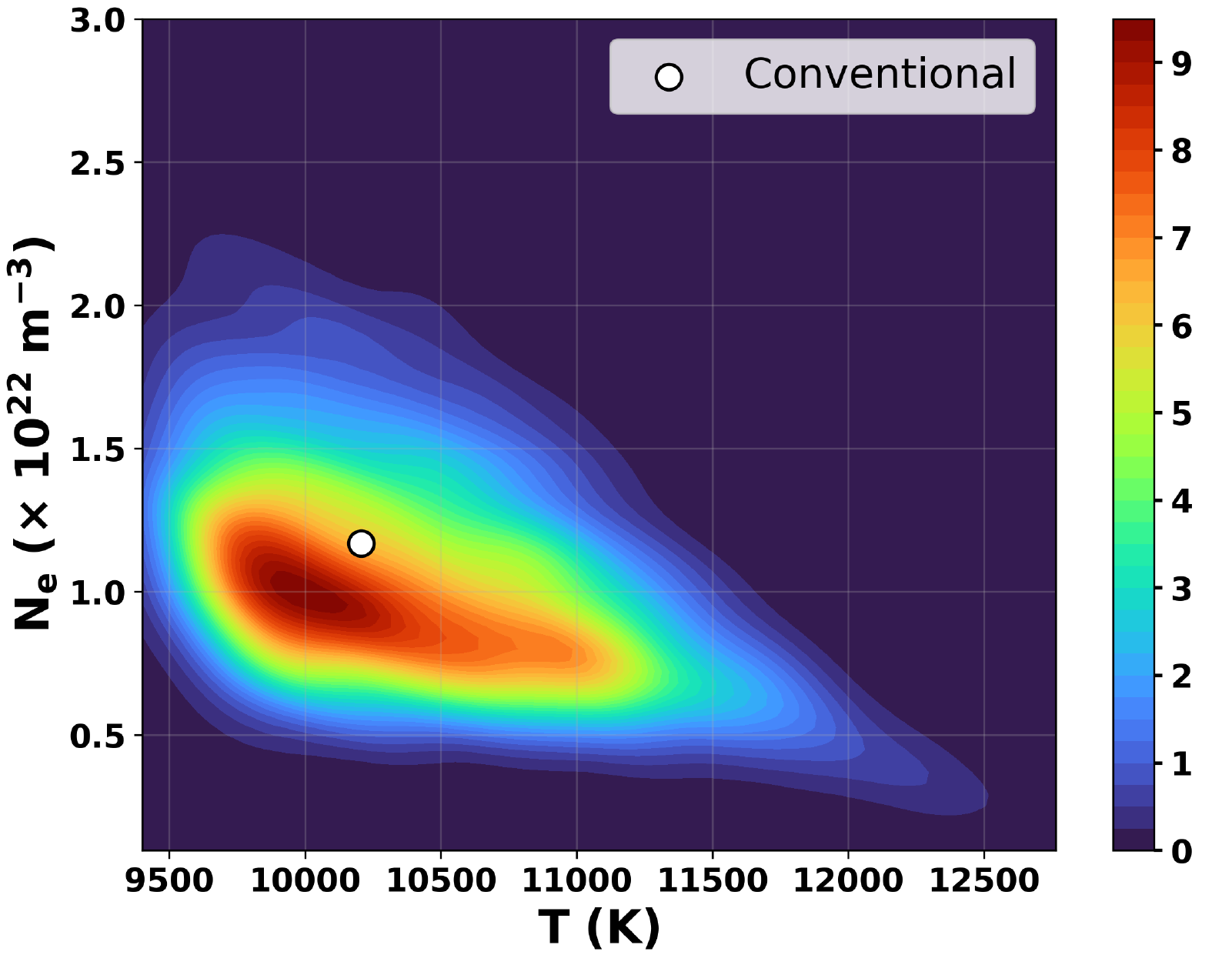}
    \caption{$T$ and $N_e$.}
    \label{fig:fig4a}
\end{subfigure}
\hfill
\begin{subfigure}{0.48\textwidth}
    \centering
    \includegraphics[width=\textwidth]{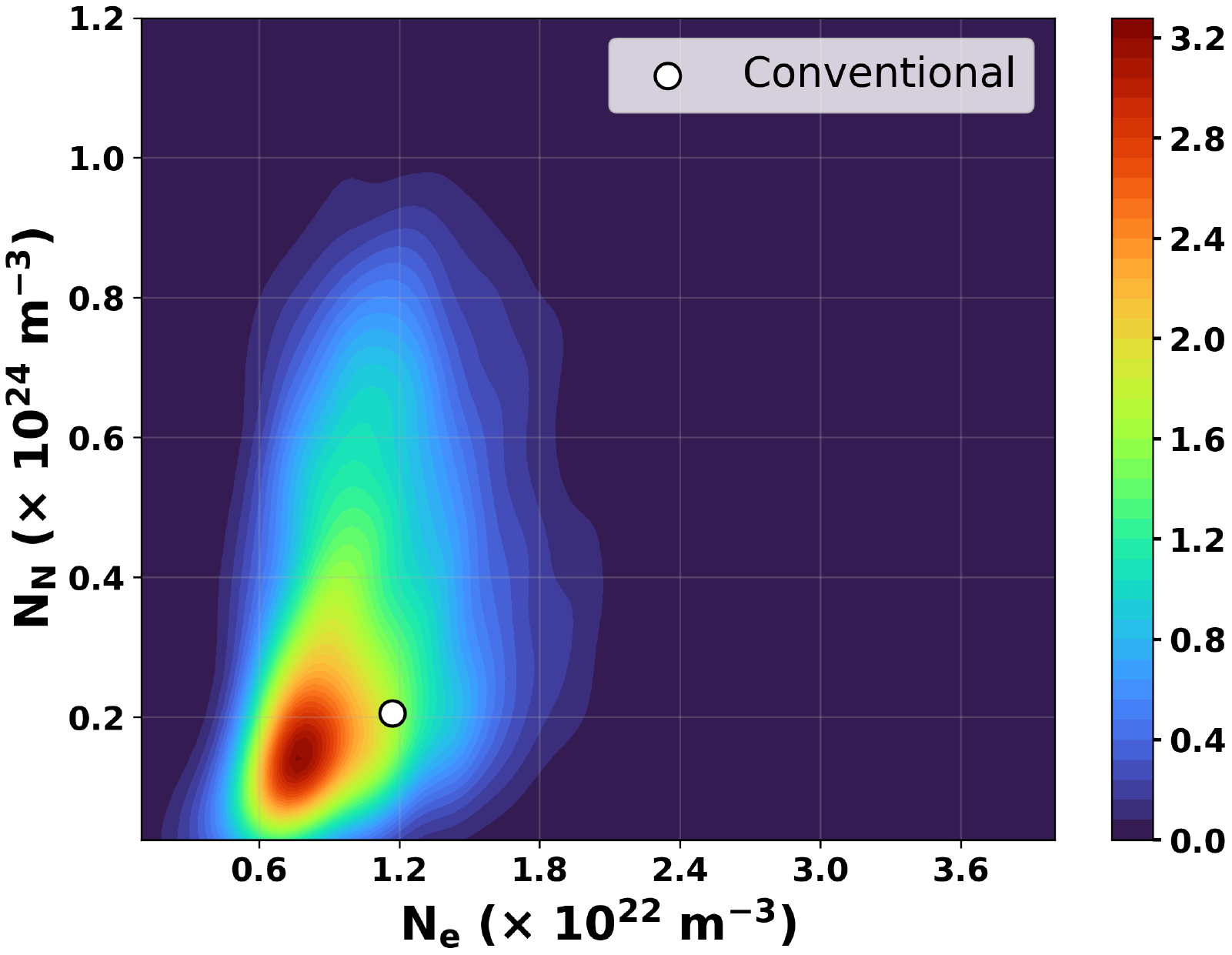}
    \caption{$N_e$ and $N_N$.}
    \label{fig:fig4b}
\end{subfigure}
\caption{Joint probability density distributions among $T$, $N_e$, and $N_N$ estimated through the MLE with the prior uncertainty in $\theta$ for the EAST T62-21.}
\label{fig:fig4}
\end{figure}

Figure~\ref{fig:fig4} shows the estimated uncertainty distributions of the nuisance parameters $\phi$, which are $T$, $N_N$, and $N_e$, obtained via MLE for the EAST T62-21 condition by following the steps described in Sec. \ref{Manufactured Data Construction}.
As discussed, the variations in $\phi$ arise from the prior uncertainties in $\theta$. The convergence of the MLE procedure was confirmed using a total of 2,400 parameter samples.
In addition, the post-shock equilibrium flow state computed without considering the uncertainty of $\theta$ is also overlaid on the joint $T$--$N$ distributions as white markers (\emph{i.e.}, Conventional). In this conventional equilibrium calculation, the Rankine-Hugoniot jump relation is coupled with the Gibbs free energy minimization.
As shown in the figure, the conventional prediction without the parametric uncertainty is enveloped by the uncertainty range of $\phi$. It is important to note that the uncertainty of $\phi$ spans about 2000~K in temperature and roughly a factor of three in number density, driven by the uncertainty in $\theta$. A similar trend was consistently observed for the EAST T62-19. This implies that without propagating the $\theta$-induced uncertainty in $\phi$ through the model expectation (see Eq. (\ref{eqn:additive_model})), the Bayesian inference for $\theta$ may be misleading.

\begin{figure}[H]
\centering
\begin{subfigure}{0.48\textwidth}
    \centering
    \includegraphics[width=\textwidth]{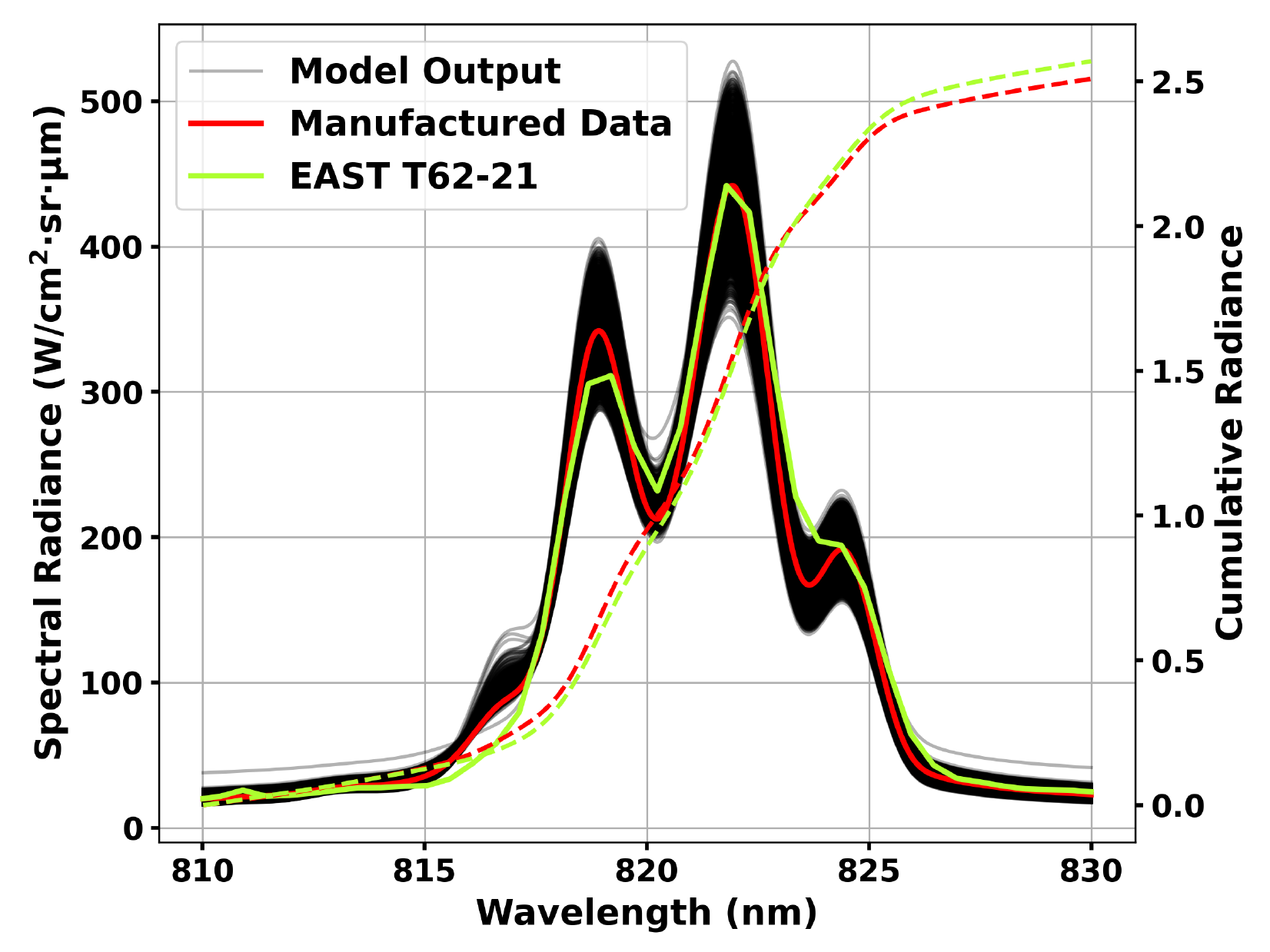}
    \caption{Variance of the spectral radiance.}
    \label{fig:fig5a}
\end{subfigure}
\hfill
\begin{subfigure}{0.48\textwidth}
    \centering
    \includegraphics[width=\textwidth]{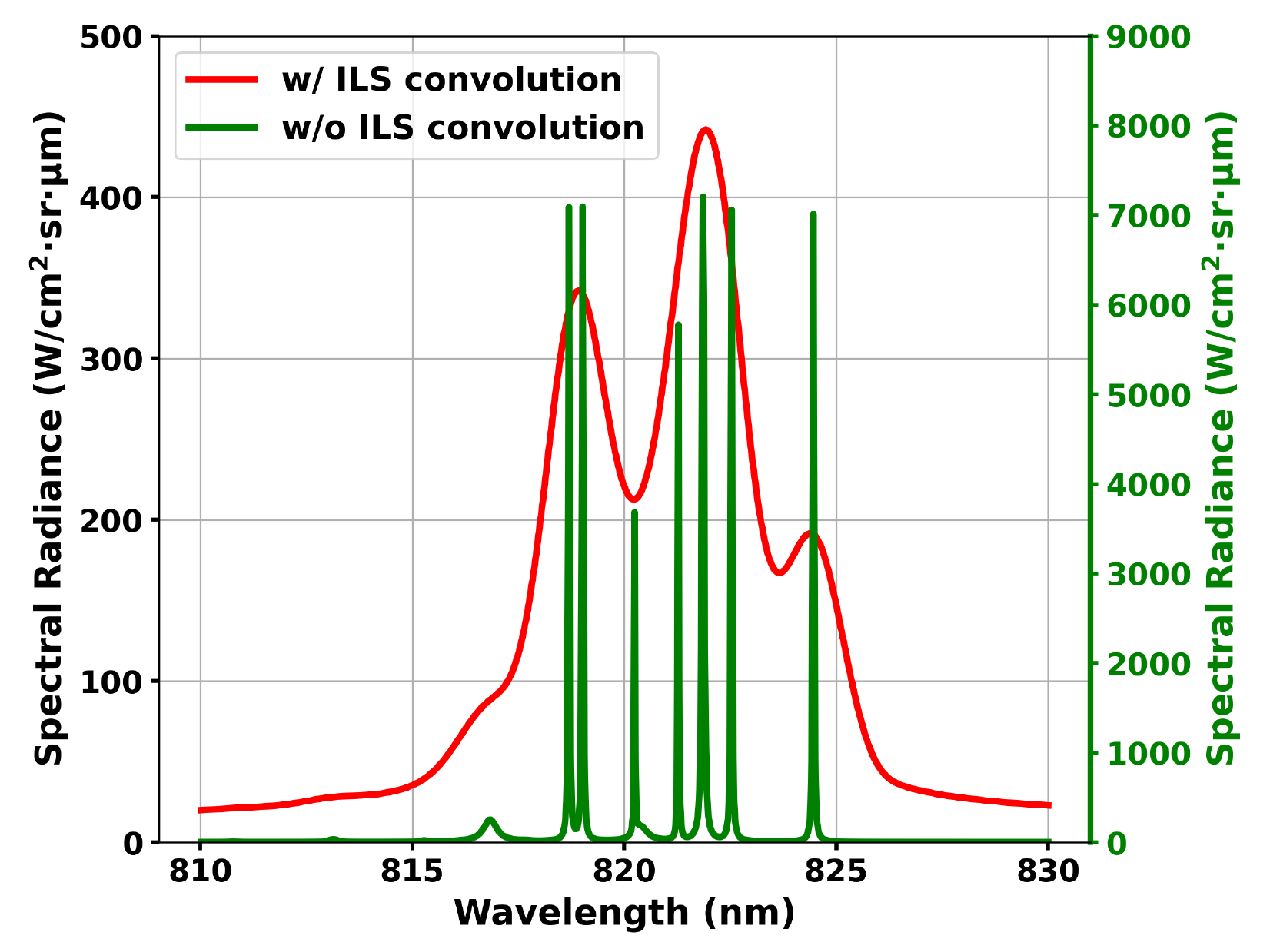}
    \caption{Manufactured data with and without the ILS convolution.}
    \label{fig:fig5b}
\end{subfigure}
\caption{Manufactured data for the Red-6 wavelength region of the EAST T62-21.}
\label{fig:fig5}
\end{figure}

Figure~\ref{fig:fig5a} shows the variance of the equilibrium spectral radiance in the Red-6 region for the shot 21 condition, obtained by propagating the distributions of $\phi$ shown in Fig. \ref{fig:fig4} to the computational model through the expectation operator. Among these results, the profile yielding the minimum SSR with the EAST measurement within the given wavelength range is selected as the manufactured data and is shown in red in the figures. Then, by eliminating the ILS convolution from the manufactured data, the individual line transitions can be resolved, as shown in Fig. \ref{fig:fig5b}, and used for Bayesian inference as $D$ in the remaining results. This procedure for constructing manufactured data was performed individually for each wavelength region considered in the present study.

\subsection{Sensitivity Analysis and Surrogate Model Performance}

\begin{figure}[h]
\centering
\begin{subfigure}{0.48\textwidth}
    \centering
    \includegraphics[width=\textwidth]{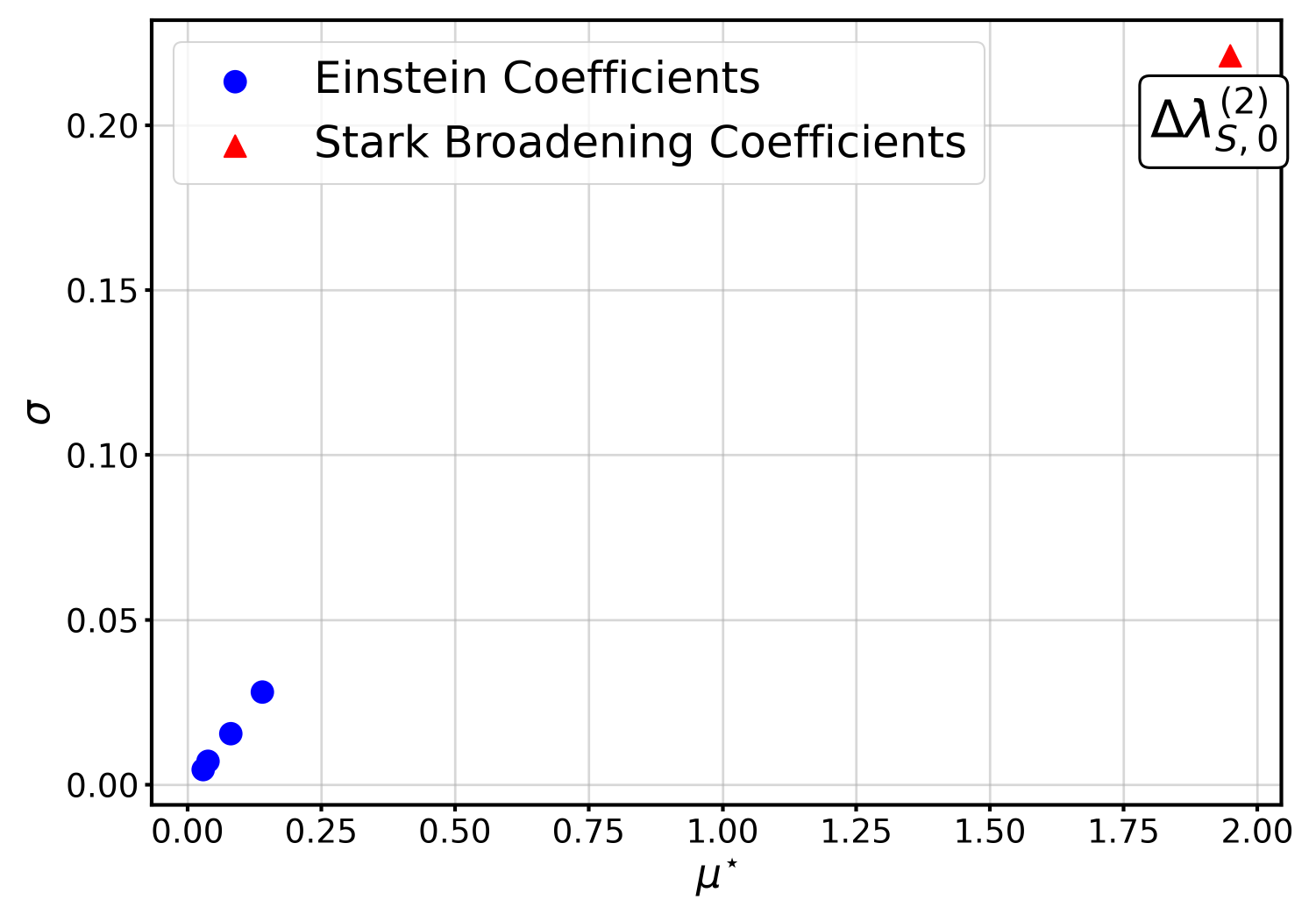}
    \caption{VUV-2}
    \label{fig:fig6a}
\end{subfigure}
\hfill
\begin{subfigure}{0.48\textwidth}
    \centering
    \includegraphics[width=\textwidth]{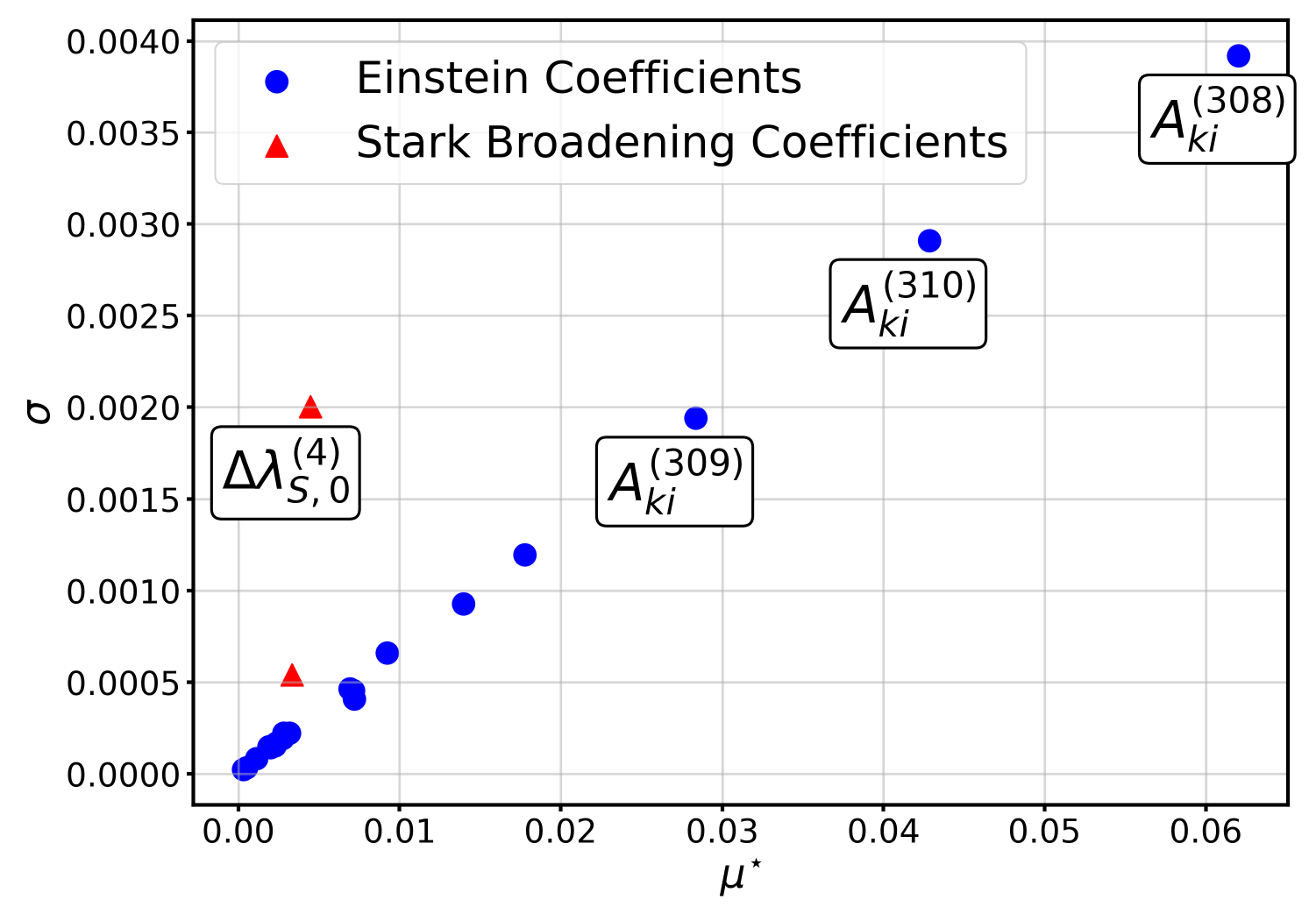}
    \caption{Red-1}
    \label{fig:fig6b}
\end{subfigure}

\vspace{0.1cm}

\begin{subfigure}{0.48\textwidth}
    \centering
    \includegraphics[width=\textwidth]{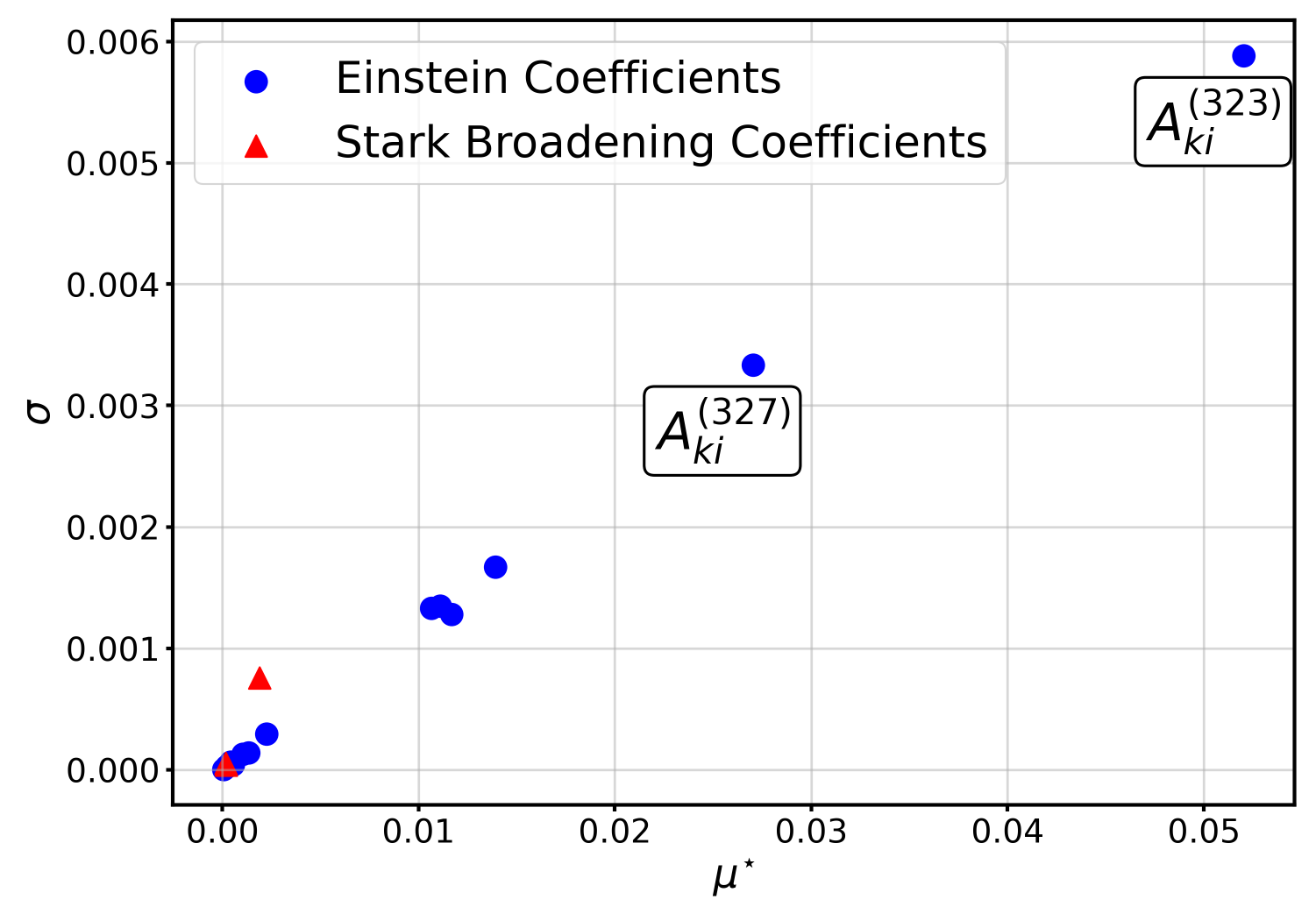}
    \caption{Red-2}
    \label{fig:fig6c}
\end{subfigure}
\hfill
\begin{subfigure}{0.48\textwidth}
    \centering
    \includegraphics[width=\textwidth]{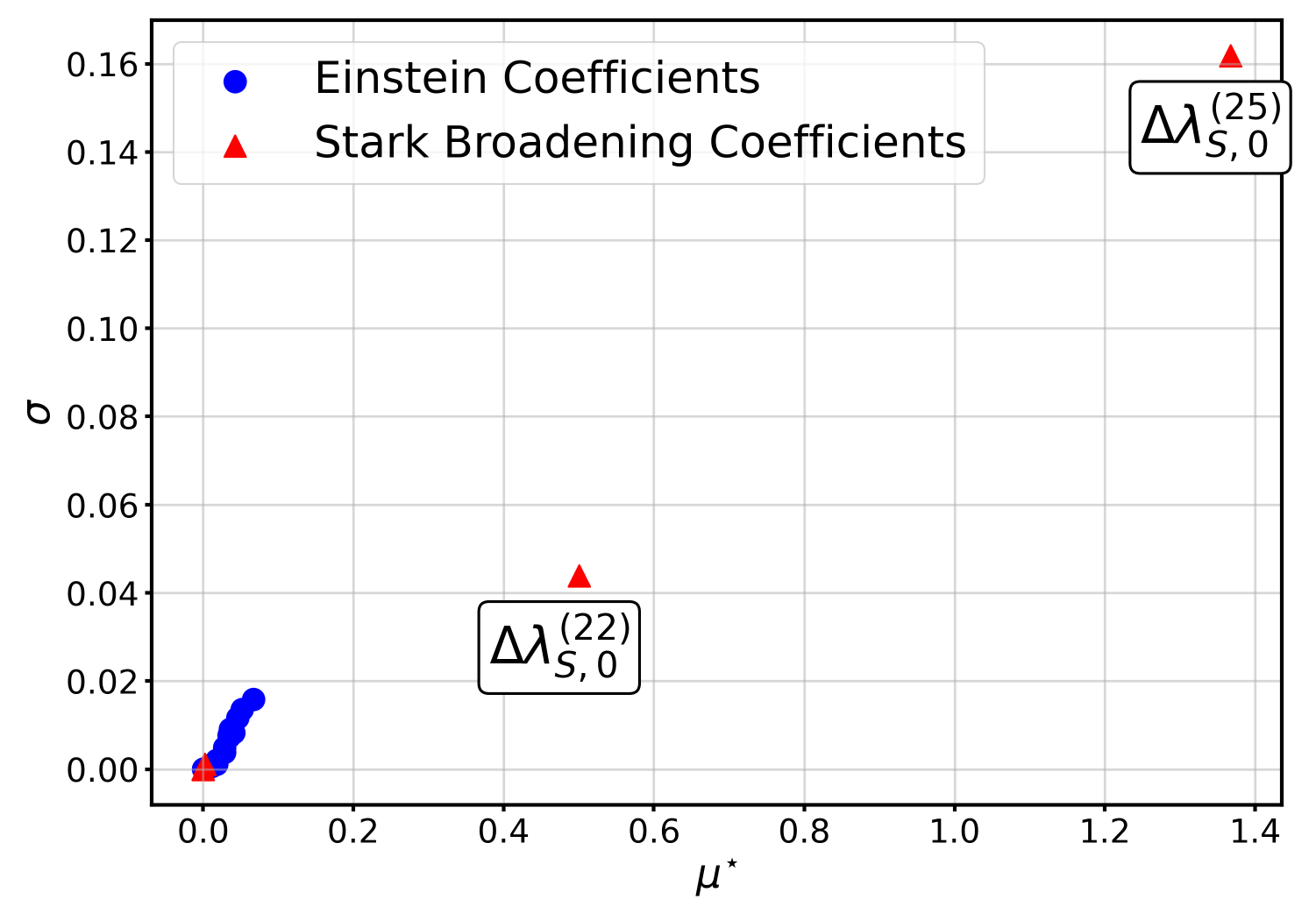}
    \caption{Red-7}
    \label{fig:fig6d}
\end{subfigure}

\caption{
Morris screening results for the EAST T62-21.
}
\label{fig:fig6}
\end{figure}

Sensitivity analyses are performed for each of the eight partitioned wavelength regions listed in Table~\ref{tab:table3} to identify the most influential parameters used for building the surrogate models and performing the Bayesian inference. In the Morris screening process, the number of trajectories used to compute the elementary effects (EE) was set to 50 for all cases.
Figure~\ref{fig:fig6} presents a subset of the Morris screening results for the EAST shot 21 condition, from which the sensitive parameters in each wavelength region can be determined. As an example, the Red-7 region shown in Fig. \ref{fig:fig6d} contains 18 Einstein coefficients and 5 Stark broadening coefficients.
Among them, $\Delta\lambda_{S,0}^{(22)}$ and $\Delta\lambda_{S,0}^{(25)}$ are identified as the sensitive parameters, because both exhibit significantly larger $\mu^{*}$ and $\sigma$ values compared with the other parameters.
It is worth noting that across all wavelength ranges, $\sigma$ is much smaller than $\mu^{*}$, indicating that the sensitive parameters influencing the integrated intensity are largely noninteracting.

After the Morris screening, as a sanity check, the variance of the model output, computed by accounting for the prior uncertainties of all parameters over the wavelength region, was compared with the variance obtained by considering only the uncertainties of the sensitive parameters. We verified that the latter case accurately reproduces the variance obtained using the full parametric uncertainties. The same procedure was applied to both EAST shots, and the overall results of the sensitivity analysis are summarized in Table~\ref{tab:table3}, which lists the influential parameters.

\begin{figure}[h]
\centering
\begin{subfigure}{0.48\textwidth}
    \centering
    \includegraphics[width=\textwidth]{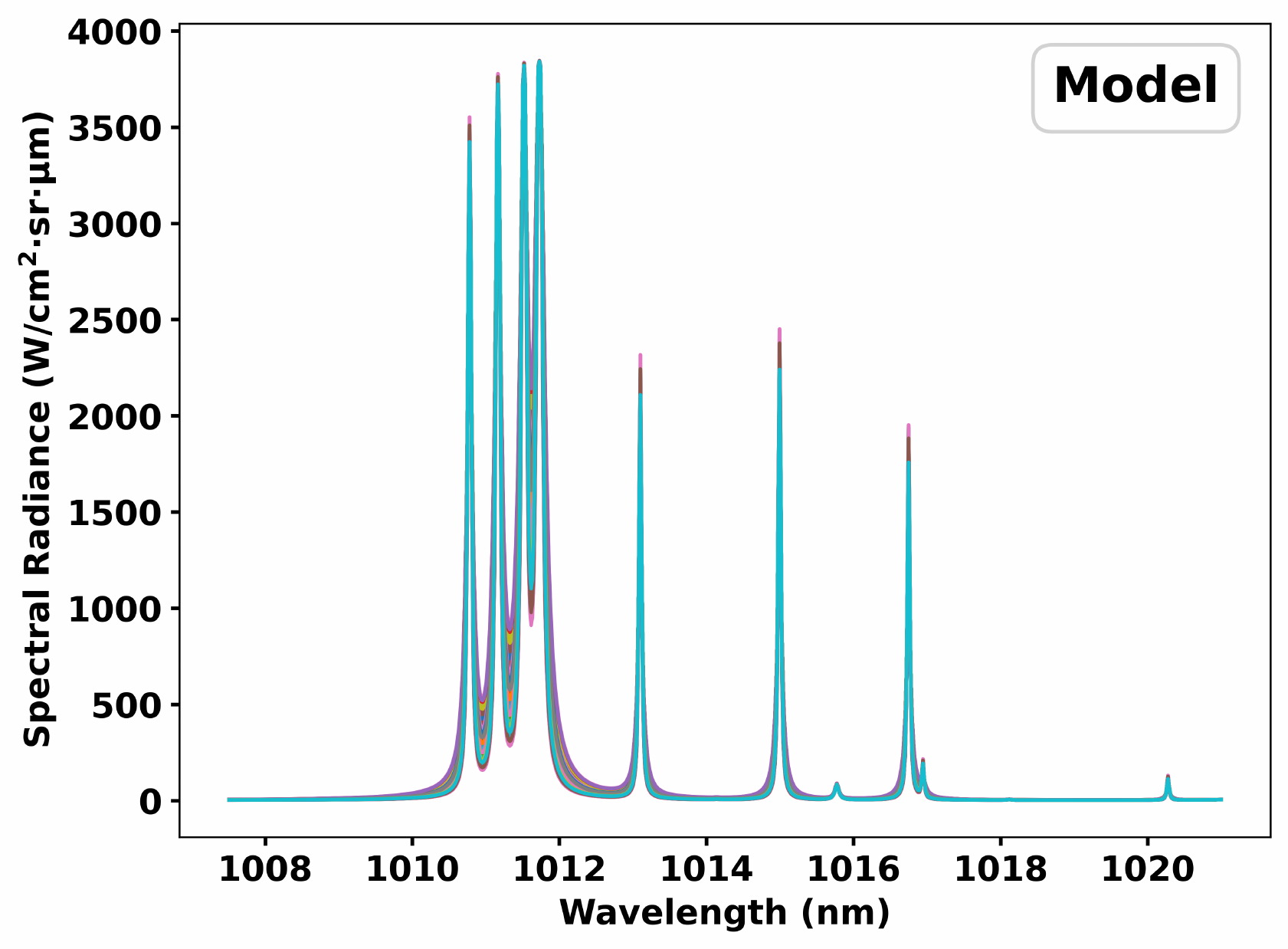}
    \caption{Computational model.}
    \label{fig:fig7a}
\end{subfigure}
\hfill
\begin{subfigure}{0.48\textwidth}
    \centering
    \includegraphics[width=\textwidth]{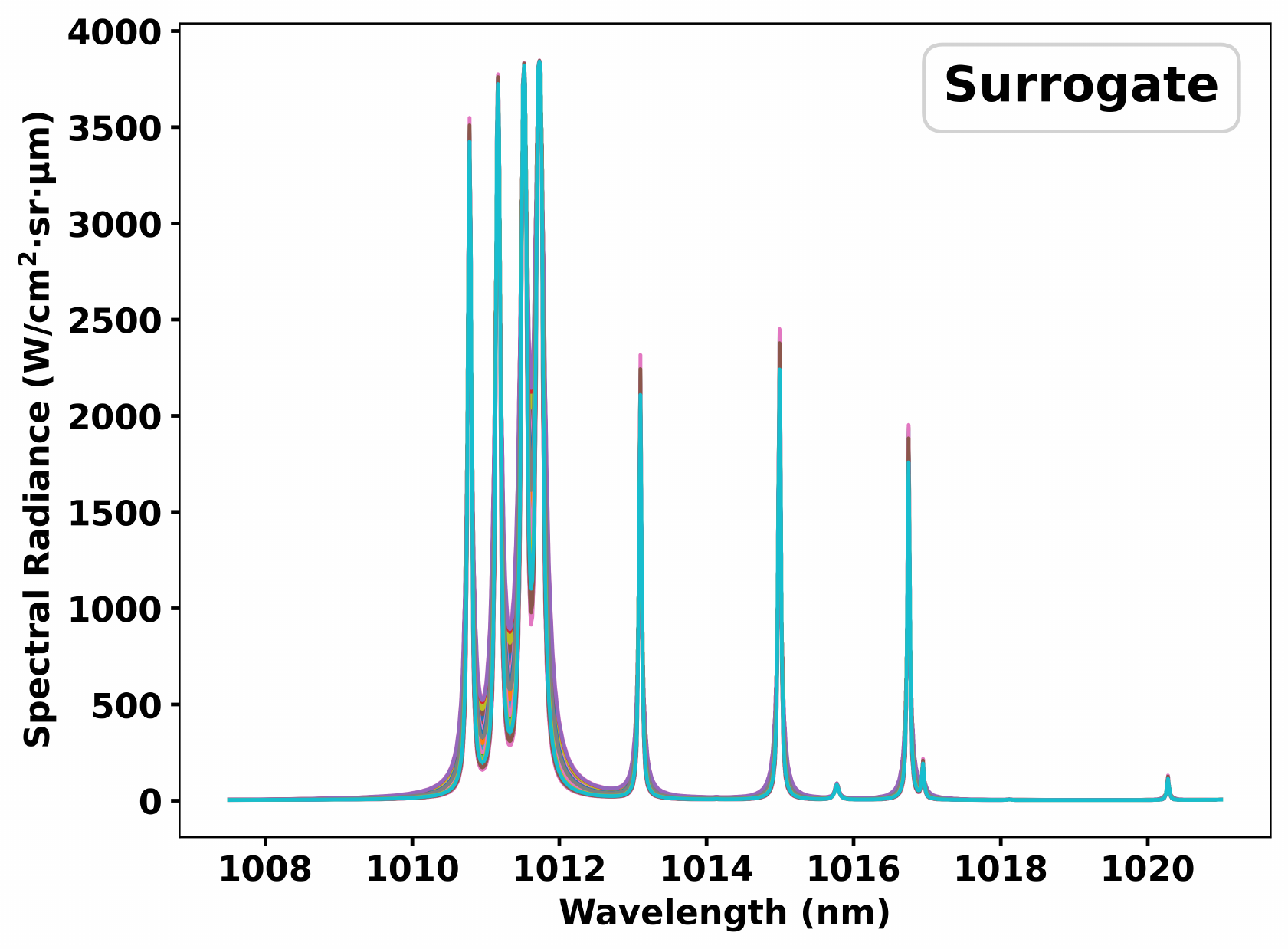}
    \caption{PCE model.}
    \label{fig:fig7b}
\end{subfigure}

\vspace{0.5em}

\begin{subfigure}{0.55\textwidth}
    \centering
    \includegraphics[width=\textwidth]{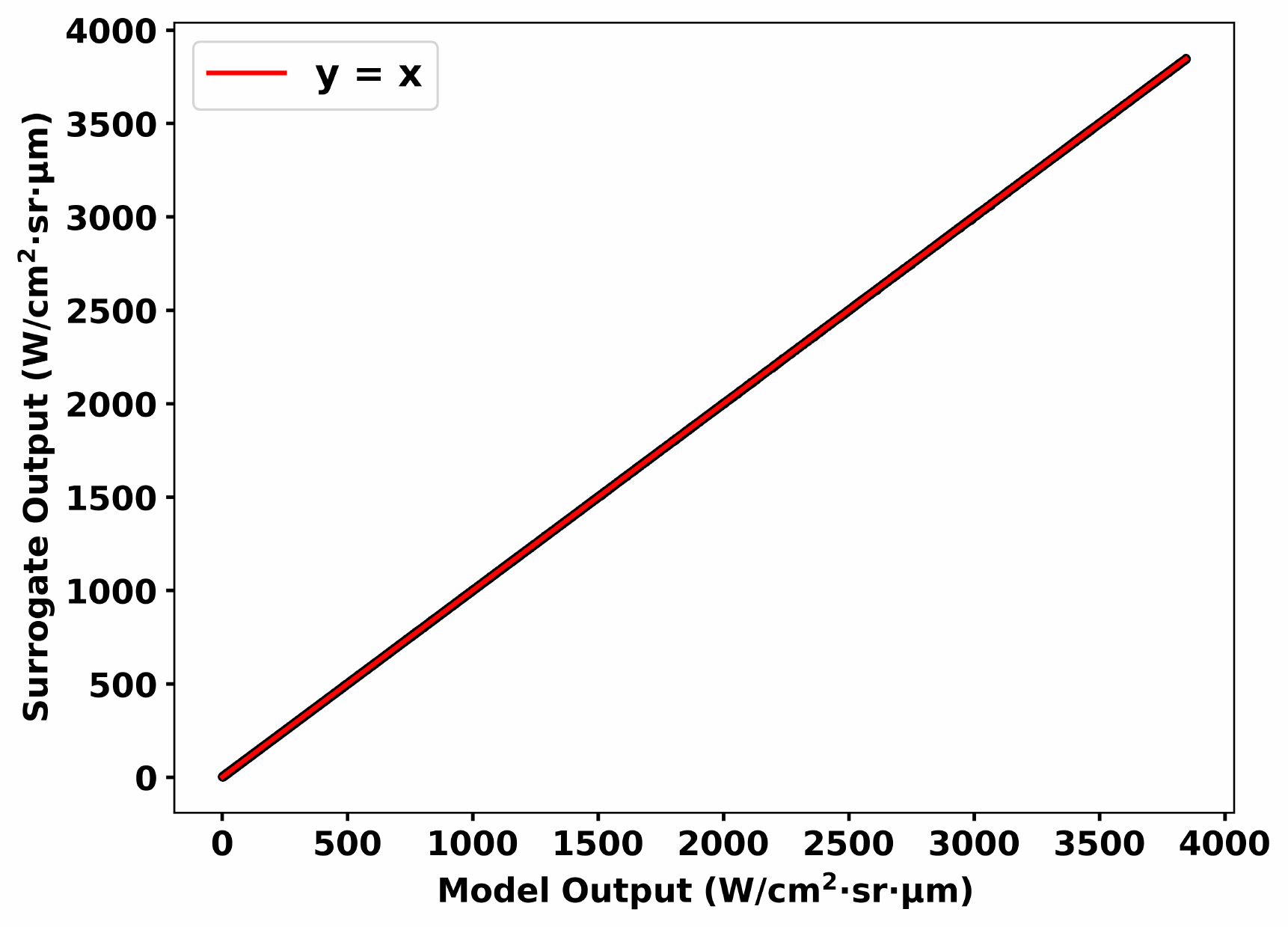}
    \caption{Parity plot.}
    \label{fig:fig7c}
\end{subfigure}

\caption{Validation of the PCE surrogate model for the IR-5 wavelength region.}
\label{fig:fig7}
\end{figure}

Next, we assessed whether the surrogate model constructed using the hybrid PCA/PCE approach accurately predicts the model expectation. Since the surrogate model output is used directly in the likelihood discrepancy term, even slightly poor predictive performance may require accounting for the model error it introduces. As a validation, the expectation computed using the surrogate model is compared with that obtained from the actual computational model, as shown in Fig.~\ref{fig:fig7} for the IR-5 region. The spectral radiance profiles predicted by the surrogate model accurately reproduce those obtained from the computational model over the wavelength range. In addition, the parity comparison shows that the surrogate outputs lie essentially on the $y=x$ line, indicating that the discrepancy between the two outputs remains close to zero across the wavelength grid points. This level of agreement was consistently observed across all wavelength regions, implying that the model error introduced by the surrogate is negligible.

\subsection{Quantified Parametric Uncertainty}

\subsubsection{VUV-2 Region}
\begin{figure}[H]
\centering
\includegraphics[width=0.7\textwidth]{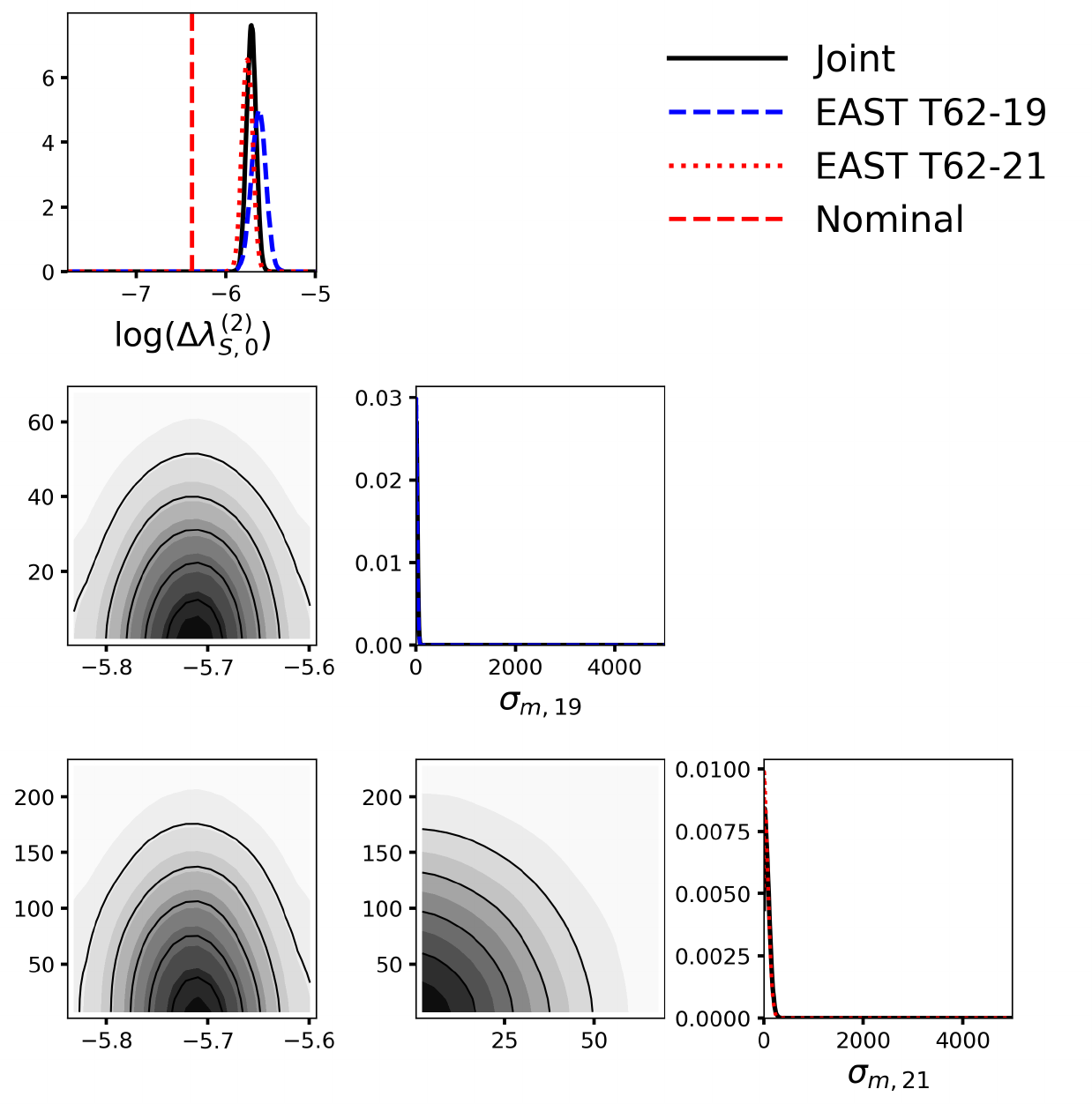}
\caption{Posterior distributions and correlations of $\Delta\lambda_{S,0}^{(2)}$, $\sigma_{m,19}$, and $\sigma_{m,21}$ in the VUV-2 region.}
\label{fig:fig8}
\end{figure}

Figure~\ref{fig:fig8} shows the posterior distribution of the Stark broadening coefficient $\Delta\lambda_{S,0}^{(2)}$ in the VUV-2 region, with the red dashed vertical line indicating the nominal value. The blue dashed line represents the distribution inferred using only the shot~19 condition, while the red dotted line corresponds to the distribution inferred using only the shot~21 condition. The black solid line shows the joint inference result obtained by considering both conditions. In all three cases, the posterior distributions of $\Delta\lambda_{S,0}^{(2)}$ exhibit similar characteristics.
Moreover, these inferred distributions have higher mean values than the nominal value, with small standard deviations. The standard deviation of the posterior distribution depends on the noise parameter $\sigma_n$ in the likelihood function and the tempering parameter $w$. If the variance of the manufactured data is large, resulting in a large $\sigma_n$, relatively larger discrepancies are allowed during the MCMC sampling process, which can lead to a broader posterior distribution. In contrast, a smaller $\sigma_n$ restricts the range of acceptable errors, resulting in a narrower posterior distribution. In addition, when the discrepancy between the MAP estimates obtained from the two shot conditions is large, yielding a smaller $w$, the influence of the data is reduced. Consequently, larger discrepancies are permitted during the MCMC sampling process, which can also lead to a broader posterior distribution. However, for $\Delta\lambda_{S,0}^{(2)}$, the distance between the MAP estimates obtained from the shot~19 and shot~21 conditions is small, yielding $w$ close to unity (0.9539). As a result, the tempering effect has only a minor influence on the width of the posterior distribution obtained from the joint inference.

Figure~\ref{fig:fig8} also presents the posterior distributions of $\sigma_{m,19}$ and $\sigma_{m,21}$, both of which are observed to be sampled near zero. This is partially due to the use of model-generated manufactured data, which leads to small discrepancies at each wavelength point. In contrast, if the EAST spectral radiance were used without explicitly accounting for measurement noise, larger values of $\sigma_m$ could be sampled to compensate for the increased discrepancies, resulting in experimental noise-driven model calibration.

\begin{figure}[H]
\centering
\begin{subfigure}{0.48\textwidth}
    \centering
    \includegraphics[width=\textwidth]{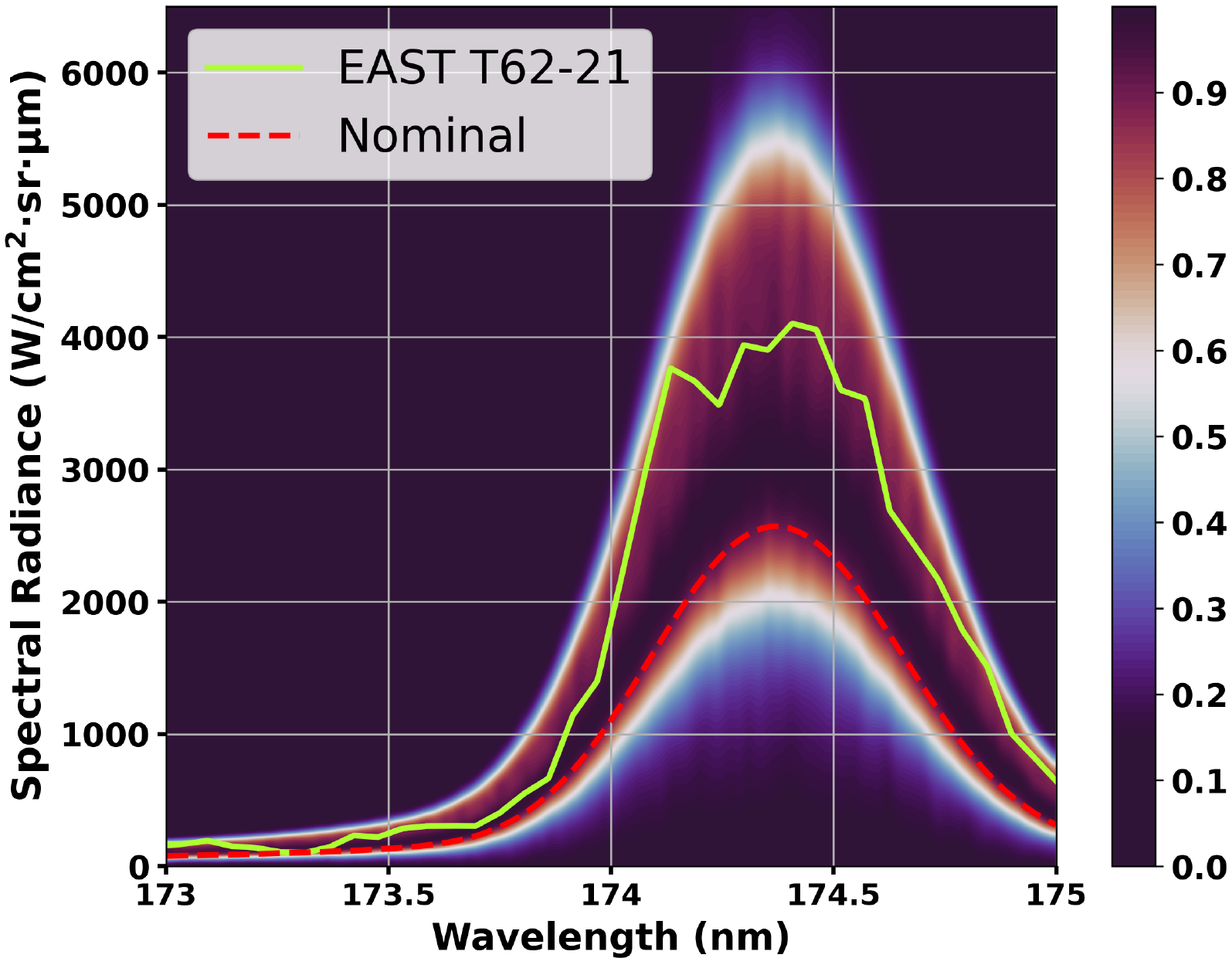}
    \caption{Contours using prior distributions of $\theta$}
    \label{fig:fig9a}
\end{subfigure}
\hfill
\begin{subfigure}{0.48\textwidth}
    \centering
    \includegraphics[width=\textwidth]{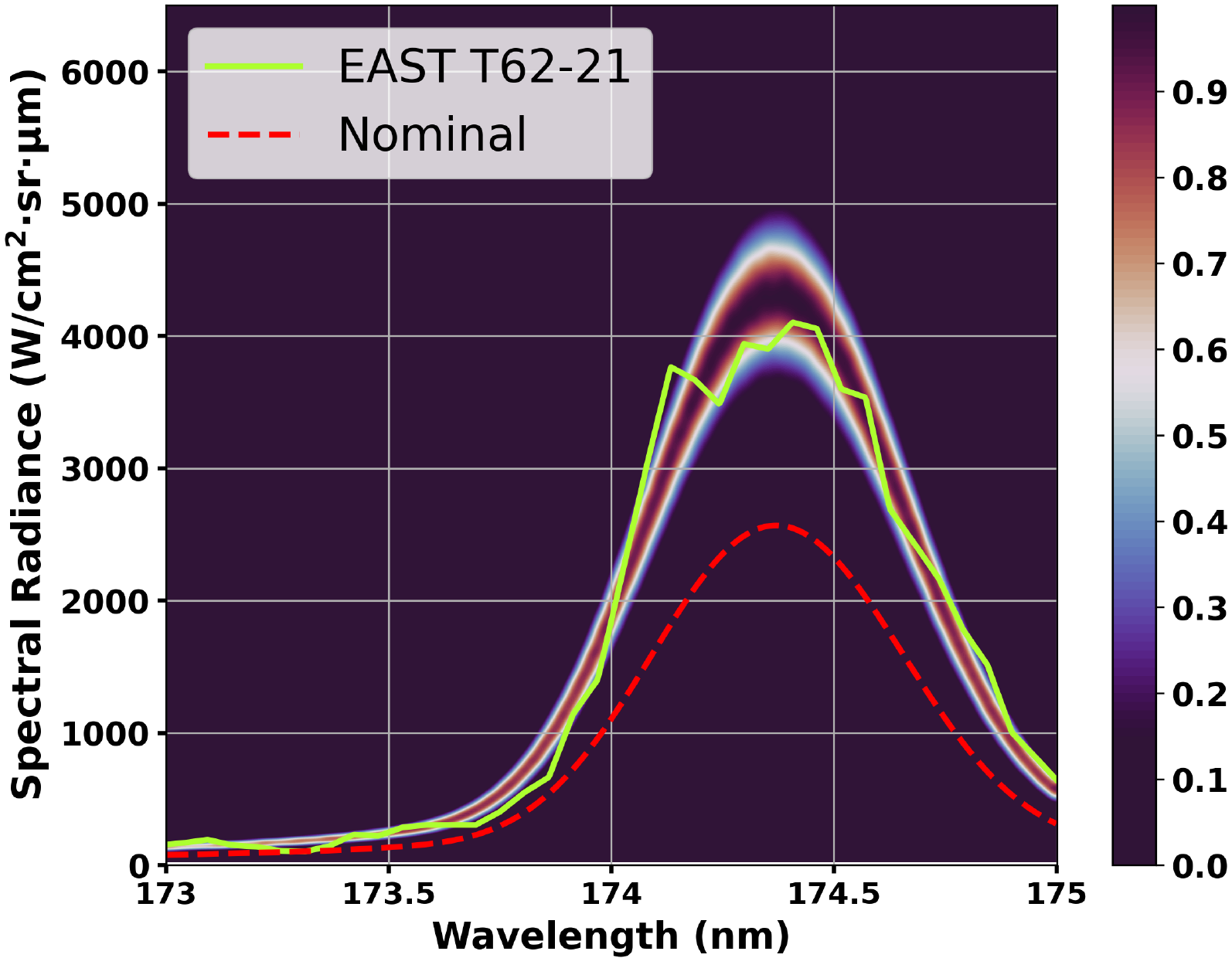}
    \caption{Contours using posterior distributions of $\theta$}
    \label{fig:fig9b}
\end{subfigure}
\caption{Model expectation contours in the VUV-2 region under the EAST T62-21.}
\label{fig:fig9}
\end{figure}

Figure~\ref{fig:fig9a} shows the contours obtained by propagating the prior distribution of $\Delta\lambda_{S,0}^{(2)}$ in the VUV-2 region through the model expectation for the shot 21 condition. Figure~\ref{fig:fig9b} presents the counterpart obtained by propagating the joint posterior distribution of $\Delta\lambda_{S,0}^{(2)}$.
The results demonstrate a significant reduction in the variance of the model expectation when using the joint posterior samples compared with the prior. It is important to note that the nominal parameter fails to reproduce the EAST-measured spectral radiance because of the large prior model uncertainty of $\Delta\lambda_{S,0}^{(2)}$. This discrepancy is resolved through the Bayesian inference, as shown in Fig.~\ref{fig:fig9b}, where the posterior contours envelop the measured profile.

\subsubsection{Red-4 Region}
\begin{figure}[H]
\centering
\includegraphics[width=0.7\textwidth]{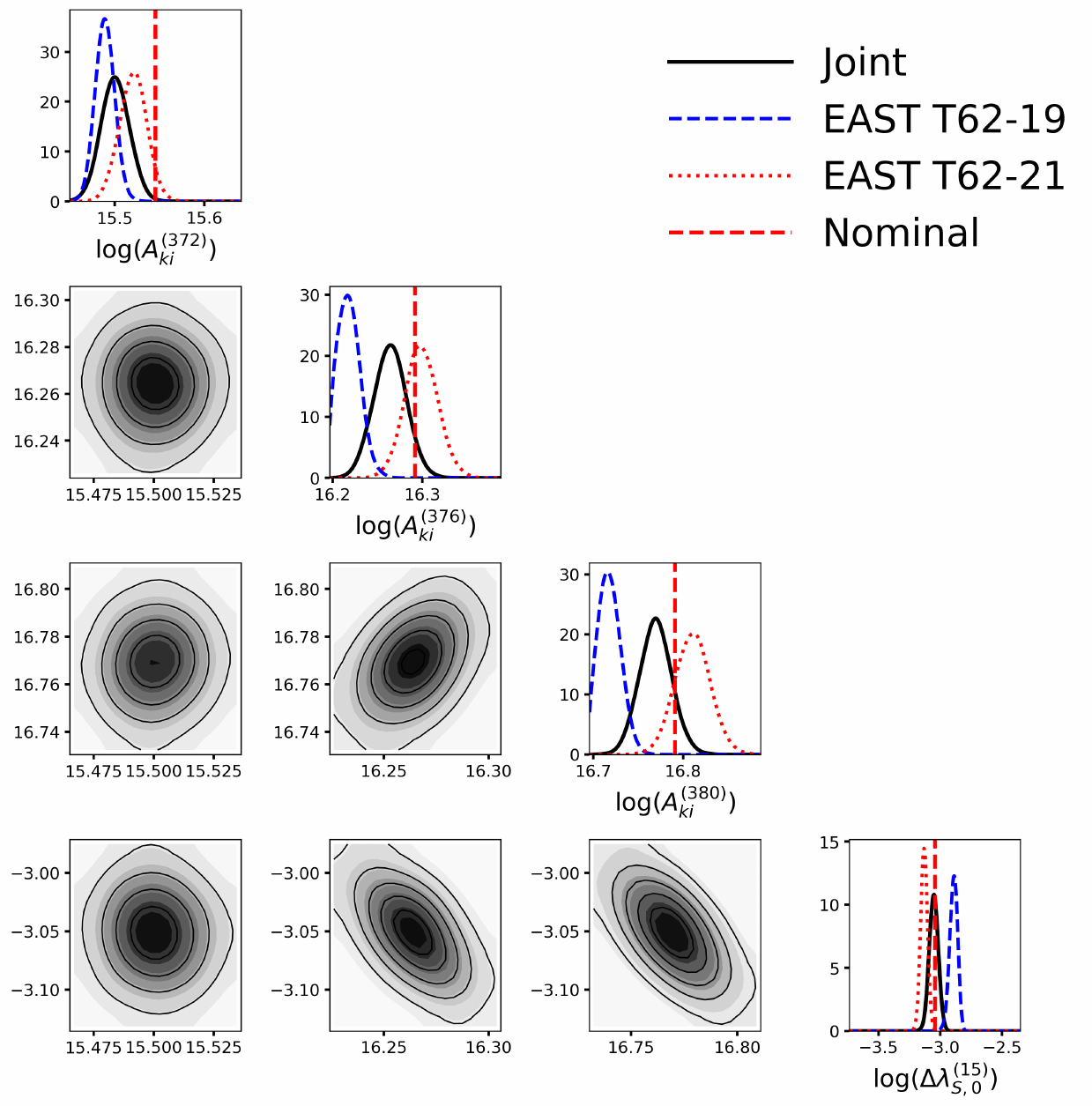}
\caption{Posterior distributions and correlations of $A_{ki}^{(372)}$, $A_{ki}^{(376)}$, $A_{ki}^{(380)}$, and $\Delta\lambda_{S,0}^{(15)}$.}
\label{fig:fig10}
\end{figure}

Figure~\ref{fig:fig10} presents the posterior distributions for the Red-4 region, which includes both Einstein coefficients and a Stark broadening coefficient as influential parameters to be inferred.
The off-diagonal elements in the figure indicate the correlations between the parameters corresponding to each row and column. These plots can show positive or negative correlations between the posterior samples of the Einstein and Stark broadening coefficients, thereby affecting error estimation at a given wavelength near the line transitions.

In the VUV-2 region, shown in Fig. \ref{fig:fig8}, the posterior distribution of $\Delta\lambda_{S,0}^{(2)}$ remains similar whether single-shot or joint conditions are considered. In contrast, as shown in Fig. \ref{fig:fig10}, this consistency is not observed for the Red-4 region. Considering only the shot 19 condition results in posterior mean values of the three Einstein coefficients that are lower than their nominal values, while the Stark broadening coefficient is overestimated. Conversely, when only the shot~21 condition is considered, the posterior means of the Einstein coefficients are close to their nominal values, but the Stark broadening coefficient is lower than the nominal value.
In this case, as mentioned earlier, the difference between the MAP estimates obtained from the two shot conditions is relatively large, yielding a comparatively low weighting parameter $w$ (0.4729). Consequently, when both conditions are considered jointly, the posterior distributions become broader than those obtained using the standard Bayesian update ($w = 1$), reflecting the reduced influence of the data. Consistent with this effect, when both conditions are considered jointly, the posterior distributions appear as a compromise between those inferred from each condition individually, rather than being dominated by a single condition.
The joint posterior tends to align closely with a specific condition if it contains more information for the inference. However, in the Red-4 region, the data from the two shots provide closely comparable information, yielding a joint posterior that reflects a balanced combination of both.

\begin{figure}[H]
\centering
\begin{subfigure}{0.48\textwidth}
    \centering
    \includegraphics[width=\textwidth]{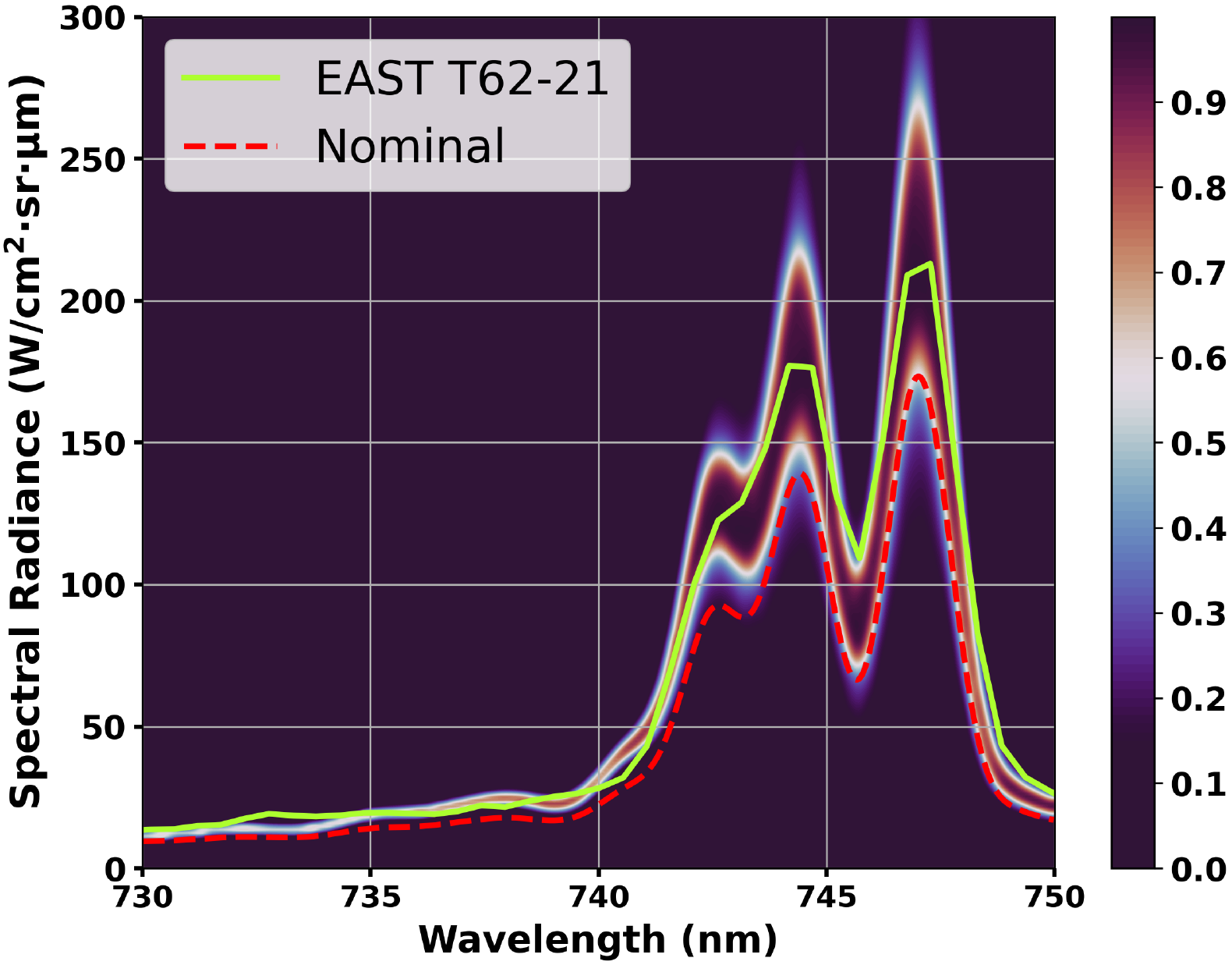}
    \caption{Contours using prior distributions of $\theta$}
    \label{fig:fig11a}
\end{subfigure}
\hfill
\begin{subfigure}{0.48\textwidth}
    \centering
    \includegraphics[width=\textwidth]{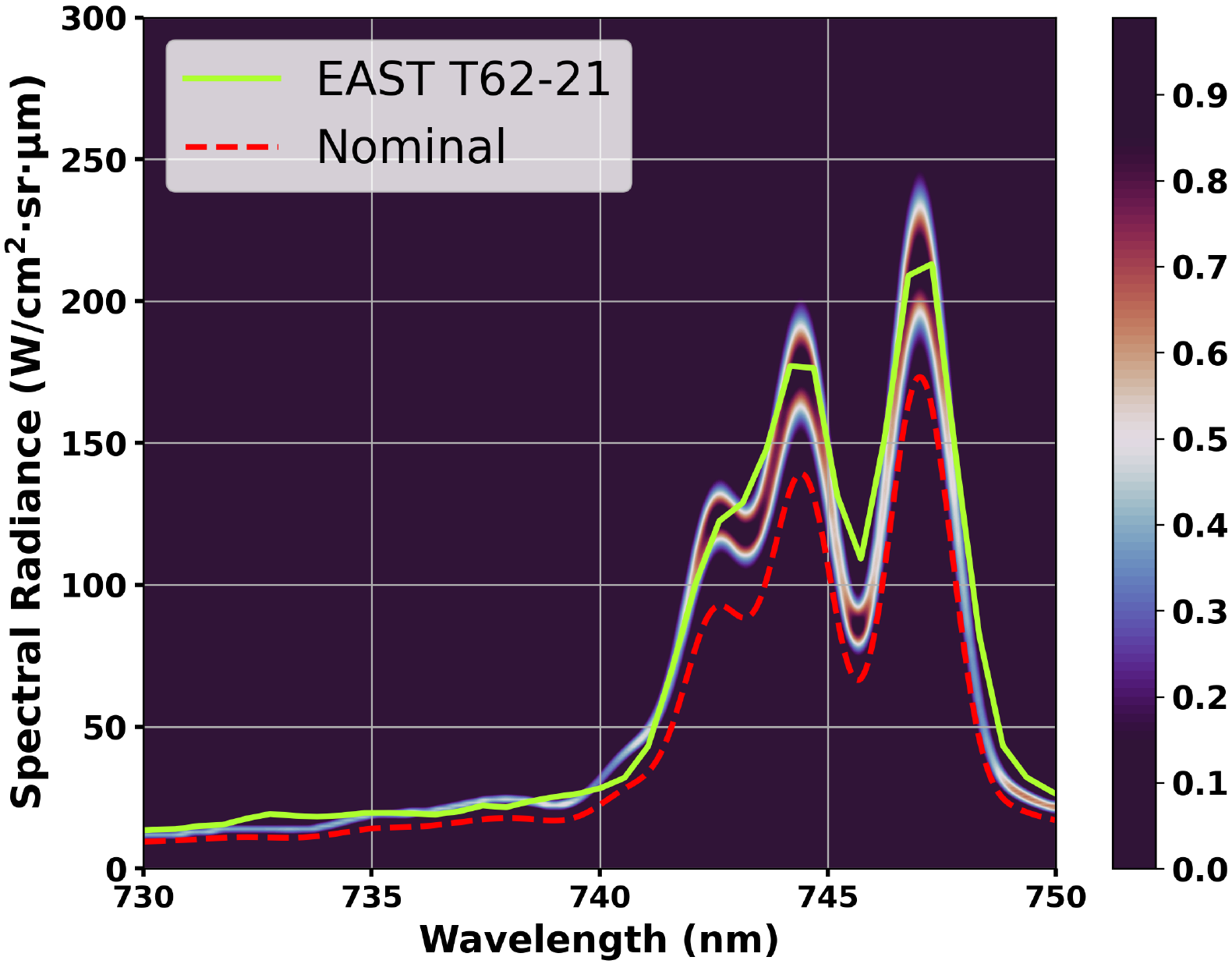}
    \caption{Contours using posterior distributions of $\theta$}
    \label{fig:fig11b}
\end{subfigure}
\caption{Model expectation contours in the Red-4 region under the EAST T62-21.}
\label{fig:fig11}
\end{figure}

Using the obtained joint posterior distributions, forward propagation through the model expectation is performed to assess the influence of the model uncertainty, as shown in Fig. \ref{fig:fig11b} for the shot 21 condition, whereas Fig. \ref{fig:fig11a} shows the counterpart with the prior distributions. It is important to note that the nominal prediction fails to reproduce the measured profile, partially due to the absence of the expectation over the nuisance parameter $\phi$, which represents the coupled uncertainty with $\theta$ as shown in Fig. \ref{fig:fig4}. With the present Bayesian inference, the uncertainty range has been significantly reduced compared to that with the prior distributions and successfully encloses the measured radiance profile. Although not shown, a similar trend has also been observed for the shot 19 case, demonstrating the robustness of the joint Bayesian inference of the present study.

\subsubsection{IR-8 Region}
\begin{figure}[H]
\centering
\includegraphics[width=0.7\textwidth]{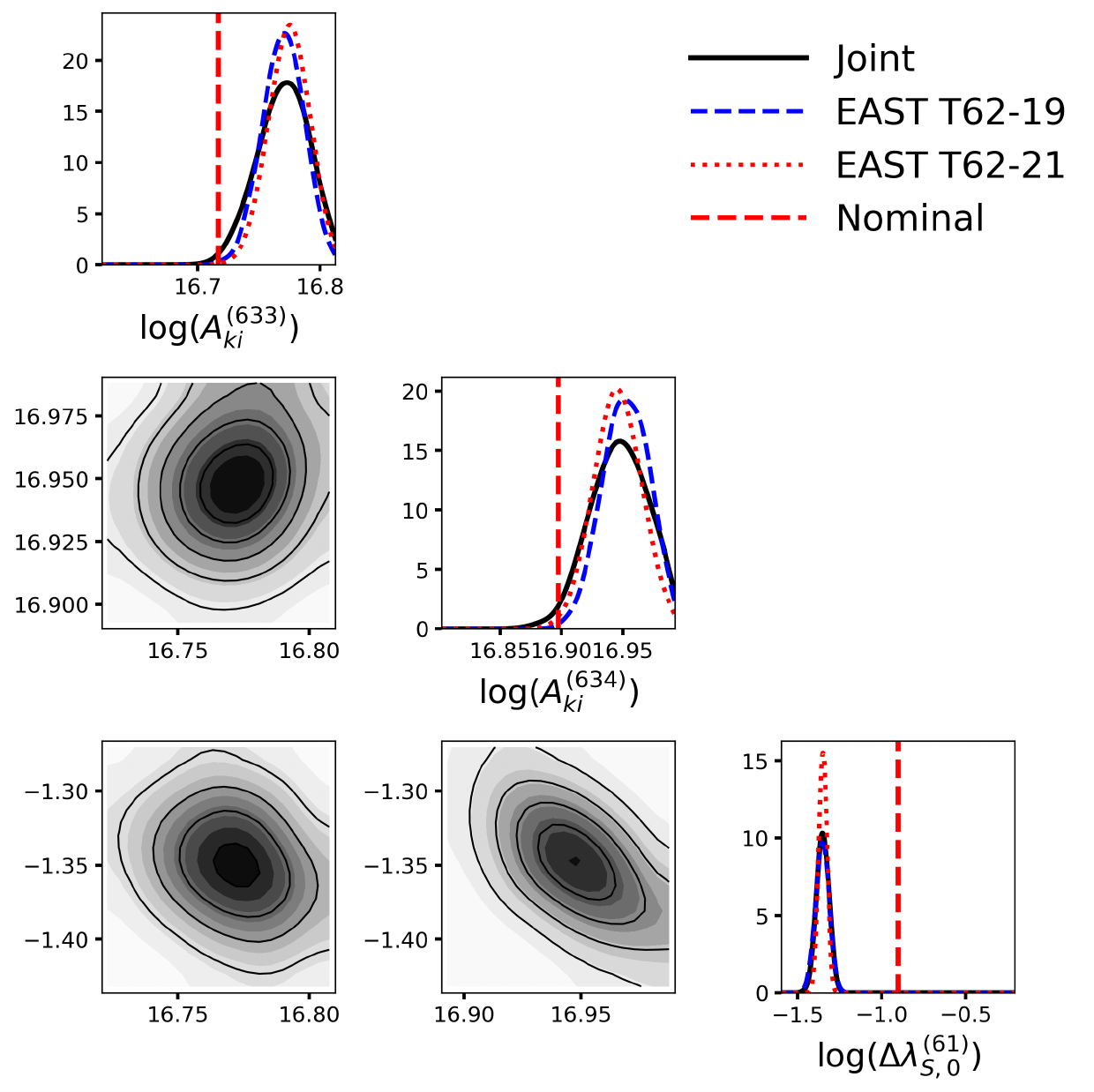}
\caption{Posterior distributions and correlations of $A_{ki}^{(633)}$, $A_{ki}^{(634)}$, and $\Delta\lambda_{S,0}^{(61)}$.}
\label{fig:fig12}
\end{figure}

Figure~\ref{fig:fig12} shows the posterior distributions for the IR-8 region, which includes two Einstein coefficients and one Stark broadening coefficient ($\Delta \lambda_{S,0}$) as the inferred parameters. The IR-8 range covers wavelengths from 1196 nm to 1264 nm, and the three parameters selected through the sensitivity analysis are associated only with the 1245--1249 nm range. Accordingly, the wavelength interval used for error estimation was restricted to the narrower range. Similar to the VUV-2 region, the posterior distributions remain similar between the single-shot calibrations and the joint estimation. This implies that the tempering parameter $w$ is computed to be close to unity, with a value of 0.9539. A negative correlation is identified between $A_{ki}^{(634)}$ and $\Delta \lambda_{S,0}^{(61)}$, given that they share the same line center. The posterior Einstein coefficients are larger than the nominal values, whereas the Stark broadening coefficient is calibrated to a smaller value than the nominal one, with this trend observed across all three inversion scenarios.

\begin{figure}[H]
\centering
\begin{subfigure}{0.48\textwidth}
    \centering
    \includegraphics[width=\textwidth]{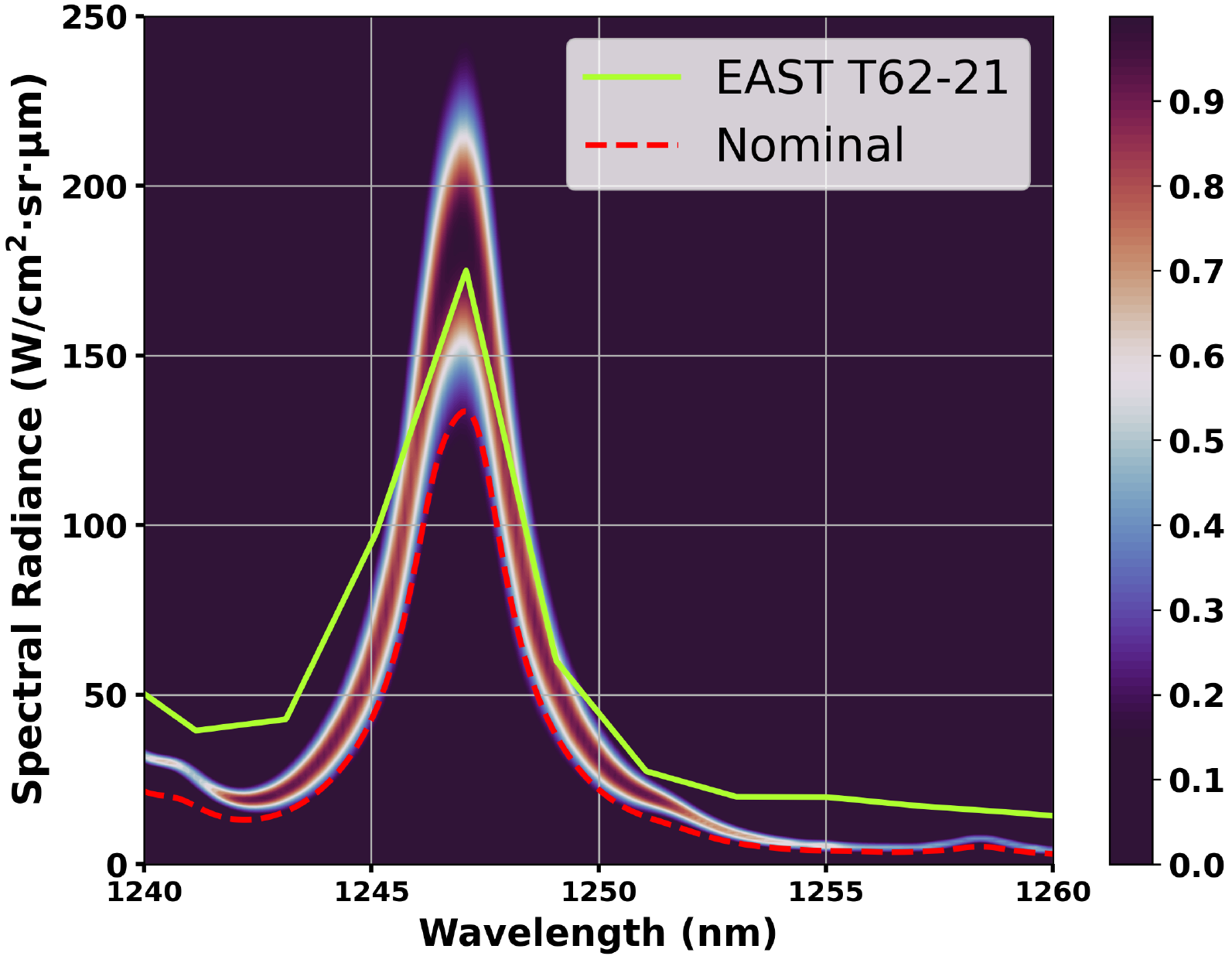}
    \caption{Contours using prior distributions of $\theta$}
    \label{fig:fig13a}
\end{subfigure}
\hfill
\begin{subfigure}{0.48\textwidth}
    \centering
    \includegraphics[width=\textwidth]{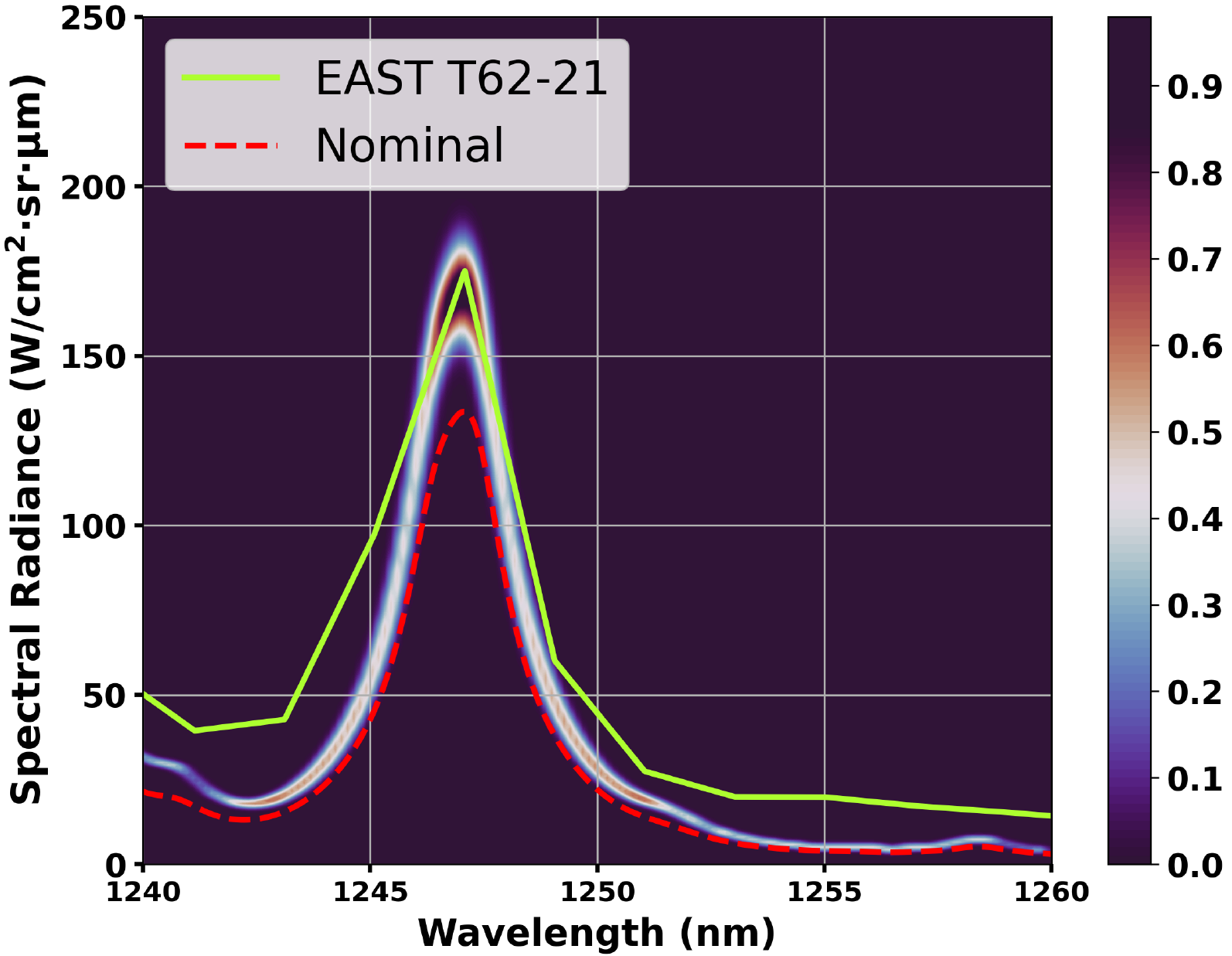}
    \caption{Contours using posterior distributions of $\theta$}
    \label{fig:fig13b}
\end{subfigure}
\caption{Model expectation contours in the IR-8 region under the EAST T62-21.}
\label{fig:fig13}
\end{figure}

The obtained joint posterior distributions are then forward propagated through the model expectation to assess the influence of the model uncertainty compared to the prior distributions, as shown in Fig.~\ref{fig:fig13}. Similar to the VUV-2 and Red-4 regions, the posterior-based result shows a significantly reduced variation in the predicted radiance compared to the prior case, especially near the peak radiance around 1247 nm. It is important to note that in the IR-8 region, the wing of the transition, lying outside the peak, is underpredicted by all models compared to the EAST measurement. These wing regions were found to be insensitive to the model expectation, which is associated with uncertainty in shock-heated temperature and species number densities. Therefore, the underprediction may be attributed to the influence of the experimental facility's background continuum radiation. This is supported by the observation that, regardless of whether bound-free and free-free continuum radiation is included in the modeling, the same underprediction persists.

In the IR-8 region, where the Einstein and Stark broadening coefficients are inferred simultaneously, the posterior distributions of the Stark broadening coefficients are found to be narrower than those of the Einstein coefficients. This can be seen from Fig. \ref{fig:fig12}, where the $x$-axis of each posterior plot spans the entire prior range. This is because, within the given prior uncertainty ranges, the Stark broadening coefficients exhibit higher sensitivity to the spectral intensity than the Einstein coefficients. Because the Stark broadening coefficients are more influential in determining the spectral radiance, they play a dominant role in the error estimation process. From the MCMC perspective, small variations in the Stark broadening coefficients can lead to large increases in the error, making such samples more likely to be rejected and resulting in the narrower posterior distribution as shown in the figure. This indicates that in the IR region, quantifying the uncertainty of the Stark broadening coefficient is more critical than that for the Einstein coefficient.

\subsubsection{Summary of Inferred Parameters}
\begin{figure}[H]
\centering
\begin{subfigure}{0.48\textwidth}
    \centering
    \includegraphics[width=\textwidth]{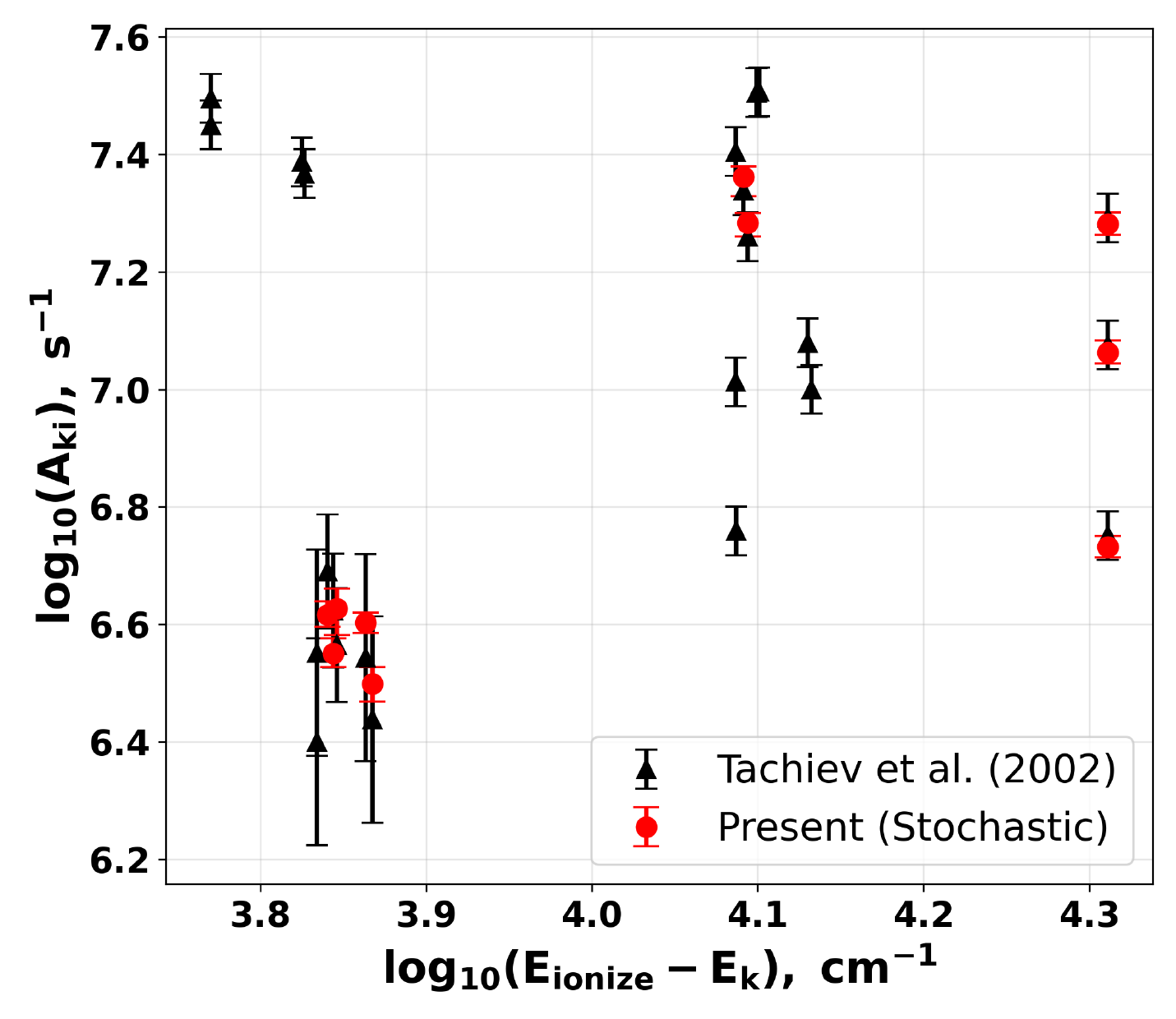}
    \caption{Einstein coefficients}
    \label{fig:fig14a}
\end{subfigure}
\hfill
\begin{subfigure}{0.48\textwidth}
    \centering
    \includegraphics[width=\textwidth]{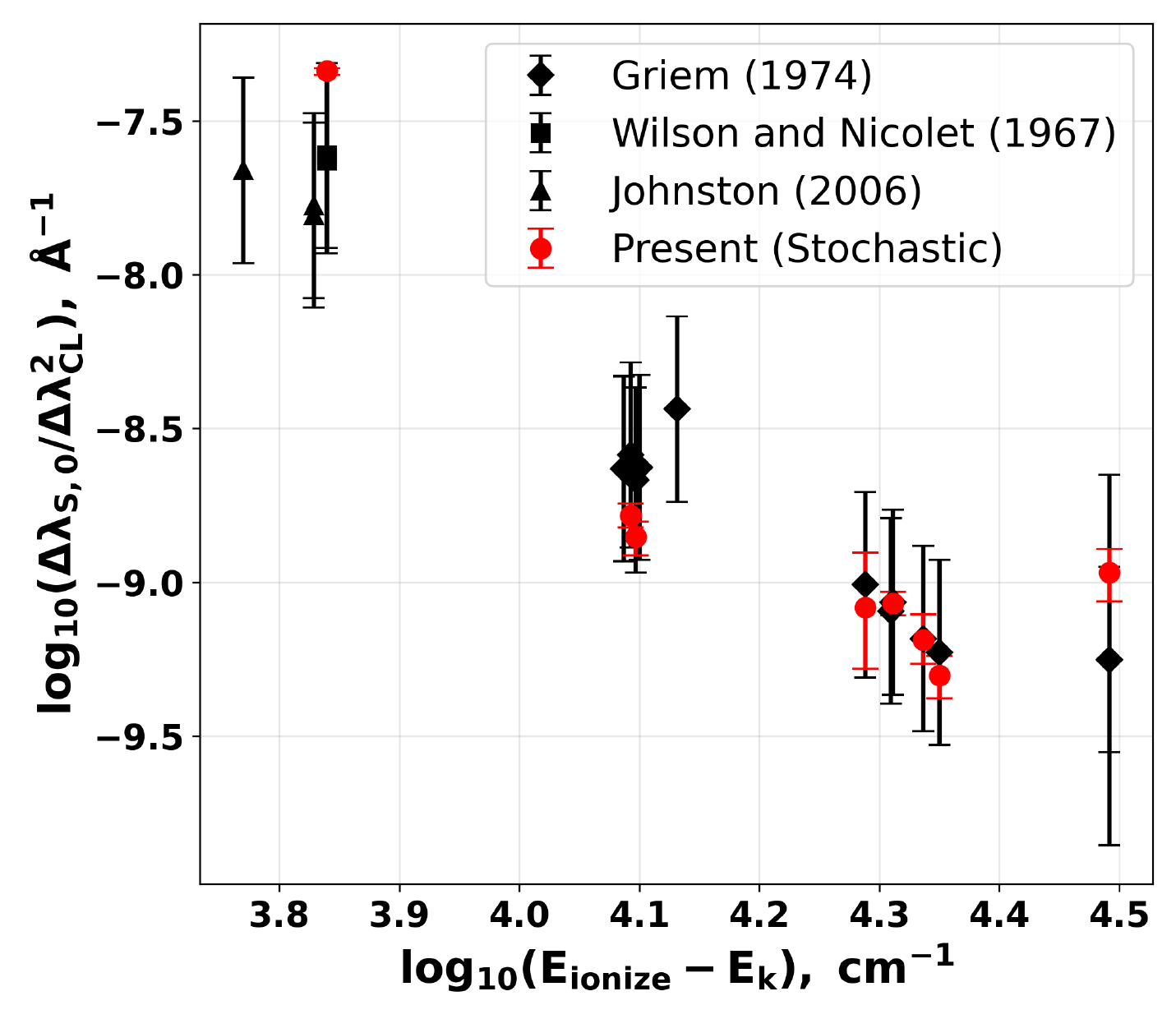}
    \caption{Stark broadening coefficients}
    \label{fig:fig14b}
\end{subfigure}
\caption{Comparison of inferred spectroscopic parameters with the literature data \cite{tachiev2002breit, 1974slbp.book.....G, wilson1967spectral, johnston2006nonequilibrium}.}
\label{fig:fig14}
\end{figure}

In the present study, the model uncertainties of 10 Einstein coefficients and 8 Stark broadening coefficients (18 parameters in total) have been inferred using the framework described in Fig. \ref{fig:fig1}.
Figure~\ref{fig:fig14} compares the prior and posterior uncertainties of these parameters, where the prior nominal values are indicated by black markers and the posterior medians by red symbols. It is important to note that the joint posterior distributions are considered more reliable than those from calibration using a single EAST shot. Therefore, hereafter, the posterior distribution refers to the joint calibration results.
As shown in the figure, the prior uncertainty ranges (\emph{i.e.}, the error bars) reported in existing studies \cite{tachiev2002breit, 1974slbp.book.....G, wilson1967spectral, johnston2006nonequilibrium} have been significantly reduced through the uncertainty quantification of the present study against the EAST data.
The resultant posterior estimates are summarized in Table~\ref{tab:table4}. The posterior median values are reported in linear space with asymmetric percentage uncertainties derived from the bounds of the 99\% credible interval (CI). The uncertainty percentages represent relative deviations from the posterior median, reported as separate lower and upper bounds to account for the skewed posterior distributions. Overall, the posterior uncertainties are significantly reduced compared to the corresponding prior ranges, indicating that the reference data provide strong constraints on the inferred parameters. The posterior parameter estimates can be used directly in radiation calculations for high-temperature atomic nitrogen, with markedly reduced model parametric uncertainty.

\begin{table}[H]
\centering
\caption{Summary of inference results for the selected spectroscopic parameters.}
\label{tab:table4}
\renewcommand{\arraystretch}{0.9}

\begin{tabular}{c c c c c}
\hline\hline
\textbf{Region} & \textbf{Symbol} 
& \textbf{Nominal value} 
& \textbf{Posterior median} 
& \textbf{Posterior uncertainty (99\% CI)} \\
\hline

% ================== VUV_2 ==================
\textbf{VUV-2} & $\Delta\lambda_{S,0}^{(2)}$
& $1.70\times10^{-3}\,\mathring{A}$
& $3.29\times10^{-3}\,\mathring{A}$
& $-13.3\% / +14.2\%$ \\
\hline
% ================== Red_1 ==================
\textbf{Red-1} & $A_{ki}^{(308)}$
& $4.90\times10^{6}\,\mathrm{s^{-1}}$
& $4.10\times10^{6}\,\mathrm{s^{-1}}$
& $-4.2\% / +7.4\%$ \\

               & $A_{ki}^{(309)}$
& $3.67\times10^{6}\,\mathrm{s^{-1}}$
& $4.21\times10^{6}\,\mathrm{s^{-1}}$
& $-14.0\% / +8.7\%$ \\

               & $A_{ki}^{(310)}$
& $4.20\times10^{6}\,\mathrm{s^{-1}}$
& $3.55\times10^{6}\,\mathrm{s^{-1}}$
& $-5.3\% / +9.5\%$ \\

               & $\Delta\lambda_{S,0}^{(4)}$
& $9.90\times10^{-1}\,\mathring{A}$
& $1.94\,\mathring{A}$
& $-4.2\% / +1.8\%$ \\

\hline

% ================== Red_2 ==================
\textbf{Red-2} & $A_{ki}^{(323)}$
& $3.49\times10^{6}\,\mathrm{s^{-1}}$
& $4.00\times10^{6}\,\mathrm{s^{-1}}$
& $-4.7\% / +5.0\%$ \\

               & $A_{ki}^{(327)}$
& $2.74\times10^{6}\,\mathrm{s^{-1}}$
& $3.15\times10^{6}\,\mathrm{s^{-1}}$
& $-8.3\% / +8.4\%$ \\

\hline

% ================== Red_4 ==================
\textbf{Red-4} & $A_{ki}^{(372)}$
& $5.64\times10^{6}\,\mathrm{s^{-1}}$
& $5.39\times10^{6}\,\mathrm{s^{-1}}$
& $-4.1\% / +4.2\%$ \\

               & $A_{ki}^{(376)}$
& $1.19\times10^{7}\,\mathrm{s^{-1}}$
& $1.15\times10^{7}\,\mathrm{s^{-1}}$
& $-4.6\% / +4.9\%$ \\

               & $A_{ki}^{(380)}$
& $1.96\times10^{7}\,\mathrm{s^{-1}}$
& $1.92\times10^{7}\,\mathrm{s^{-1}}$
& $-4.3\% / +4.7\%$ \\

               & $\Delta\lambda_{S,0}^{(15)}$
& $4.77\times10^{-2}\,\mathring{A}$
& $4.72\times10^{-2}\,\mathring{A}$
& $-9.2\% / +9.8\%$ \\

\hline

% ================== Red_6 ==================
\textbf{Red-6} & $\Delta\lambda_{S,0}^{(21)}$
& $4.42\times10^{-2}\,\mathring{A}$
& $4.37\times10^{-2}\,\mathring{A}$
& $-15.8\% / +17.7\%$ \\

\hline

% ================== Red_7 ==================
\textbf{Red-7} & $\Delta\lambda_{S,0}^{(22)}$
& $7.28\times10^{-2}\,\mathring{A}$
& $6.27\times10^{-2}\,\mathring{A}$
& $-27.2\% / +31.2\%$ \\

               & $\Delta\lambda_{S,0}^{(25)}$
& $4.48\times10^{-2}\,\mathring{A}$
& $3.78\times10^{-2}\,\mathring{A}$
& $-9.6\% / +10.3\%$ \\

\hline

% ================== IR_5 ==================
\textbf{IR-5} & $\Delta\lambda_{S,0}^{(40)}$
& $2.21\times10^{-1}\,\mathring{A}$
& $1.43\times10^{-1}\,\mathring{A}$
& $-11.7\% / +12.7\%$ \\

\hline

% ================== IR_8 ==================
\textbf{IR-8} & $A_{ki}^{(633)}$
& $1.82\times10^{7}\,\mathrm{s^{-1}}$
& $1.92\times10^{7}\,\mathrm{s^{-1}}$
& $-5.5\% / +4.1\%$ \\

               & $A_{ki}^{(634)}$
& $2.18\times10^{7}\,\mathrm{s^{-1}}$
& $2.29\times10^{7}\,\mathrm{s^{-1}}$
& $-6.6\% / +4.4\%$ \\

               & $\Delta\lambda_{S,0}^{(61)}$
& $4.06\times10^{-1}\,\mathring{A}$
& $2.59\times10^{-1}\,\mathring{A}$
& $-9.6\% / +9.9\%$ \\

\hline\hline
\end{tabular}
\end{table}

\subsubsection{Impact of Model Uncertainty on Radiative Heating}

It is important to assess how the newly updated parametric uncertainties for the spectroscopic parameters obtained in this study, together with the previously reported prior distributions, affect radiative heat flux predictions under realistic hypersonic atmospheric entry flow conditions. To this end, we consider a 3~m radius sphere entering the Earth's atmosphere at velocities of 10, 12, and 14 km/s to generate stagnation-line profiles and assess the radiative heat flux using the prior and posterior parametric uncertainty ranges. The flow field simulations account for thermochemical nonequilibrium using a two-temperature (2T) model for an 11-species air mixture \cite{park1993review}. Simulations are performed with the CFD solver \textsc{hegel} \cite{munafo2020computational,munafo2024hegel}, which uses a second-order finite-volume method to discretize the thermochemical nonequilibrium Navier-Stokes equations. The freestream temperature and pressure are set to 200 K and 5 Pa, respectively, to model a super-orbital hypersonic entry flight trajectory. An isothermal boundary condition at 500 K is imposed along the sphere surface.

To clearly assess the impact of uncertainty in the spectroscopic parameters, the atomic nitrogen bound-bound line transition is the only system of interest in the radiation calculation. The effect of non-Boltzmann electronic population distributions is accounted for to obtain realistic radiative heat flux values. The quasi-steady state (QSS) solution of the electronic state-to-state master equation was obtained using the kinetic database of Jo \emph{et al.} \cite{jo2019electronic} that includes both electron and heavy-particle impact excitation and electron impact ionization.

\begin{figure}[H]
\centering
\begin{subfigure}{0.55\textwidth}
    \centering
    \includegraphics[width=\textwidth]{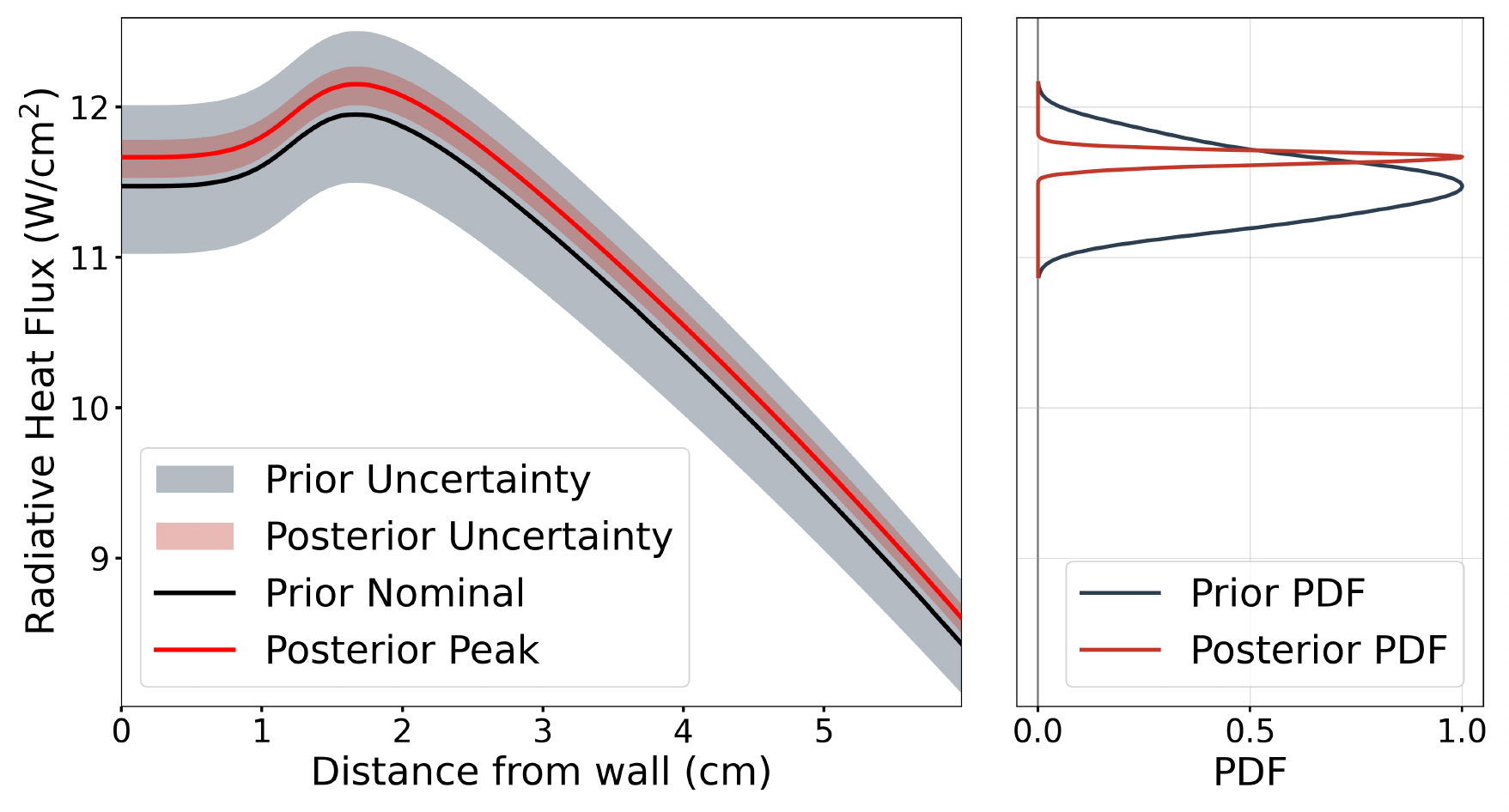}
    \caption{10 km/s.}
    \label{fig:fig15a}
\end{subfigure}
\medskip
\begin{subfigure}{0.55\textwidth}
    \centering
    \includegraphics[width=\textwidth]{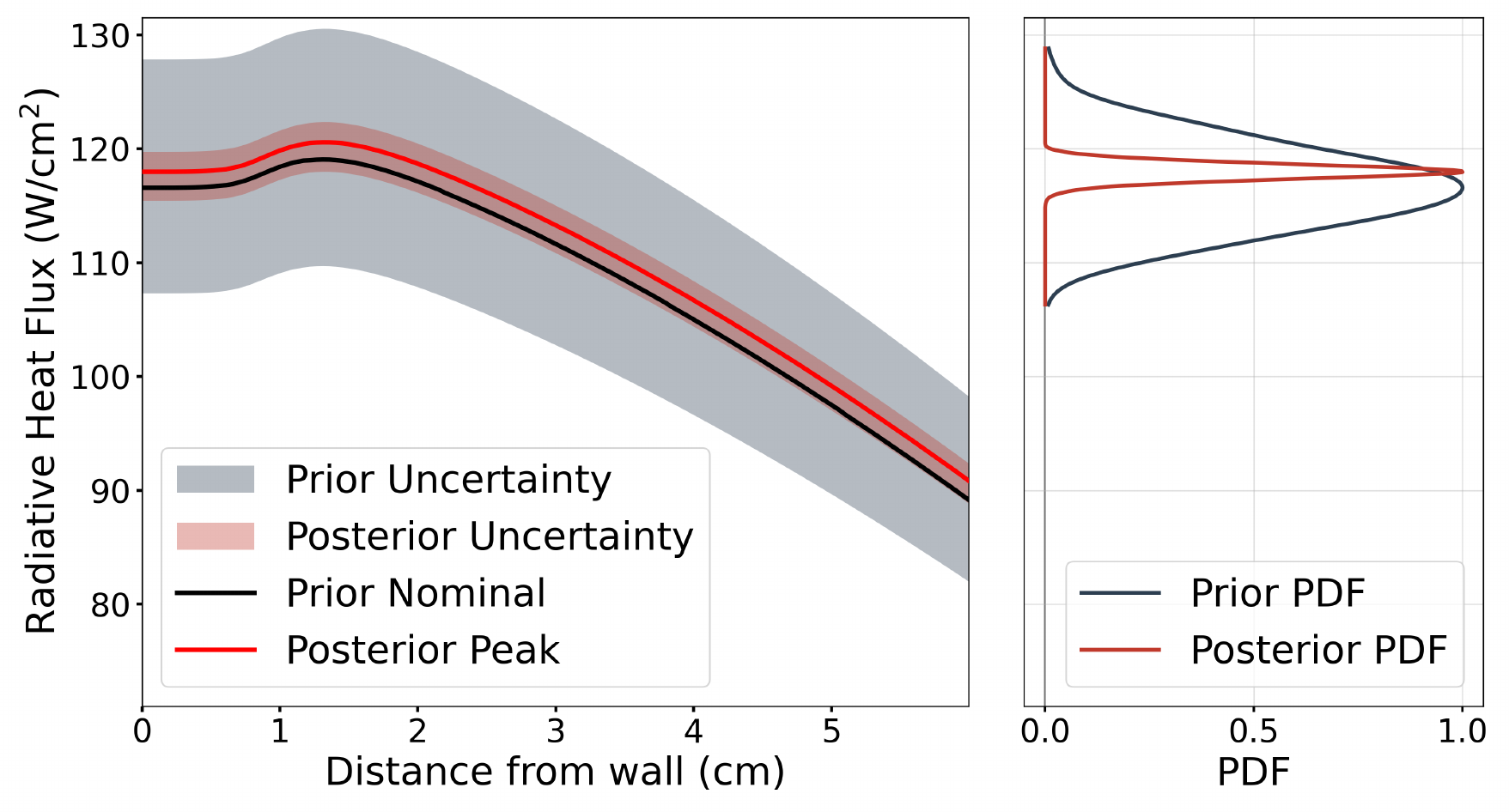}
    \caption{12 km/s.}
    \label{fig:fig15b}
\end{subfigure}
\medskip
\begin{subfigure}{0.55\textwidth}
    \centering
    \includegraphics[width=\textwidth]{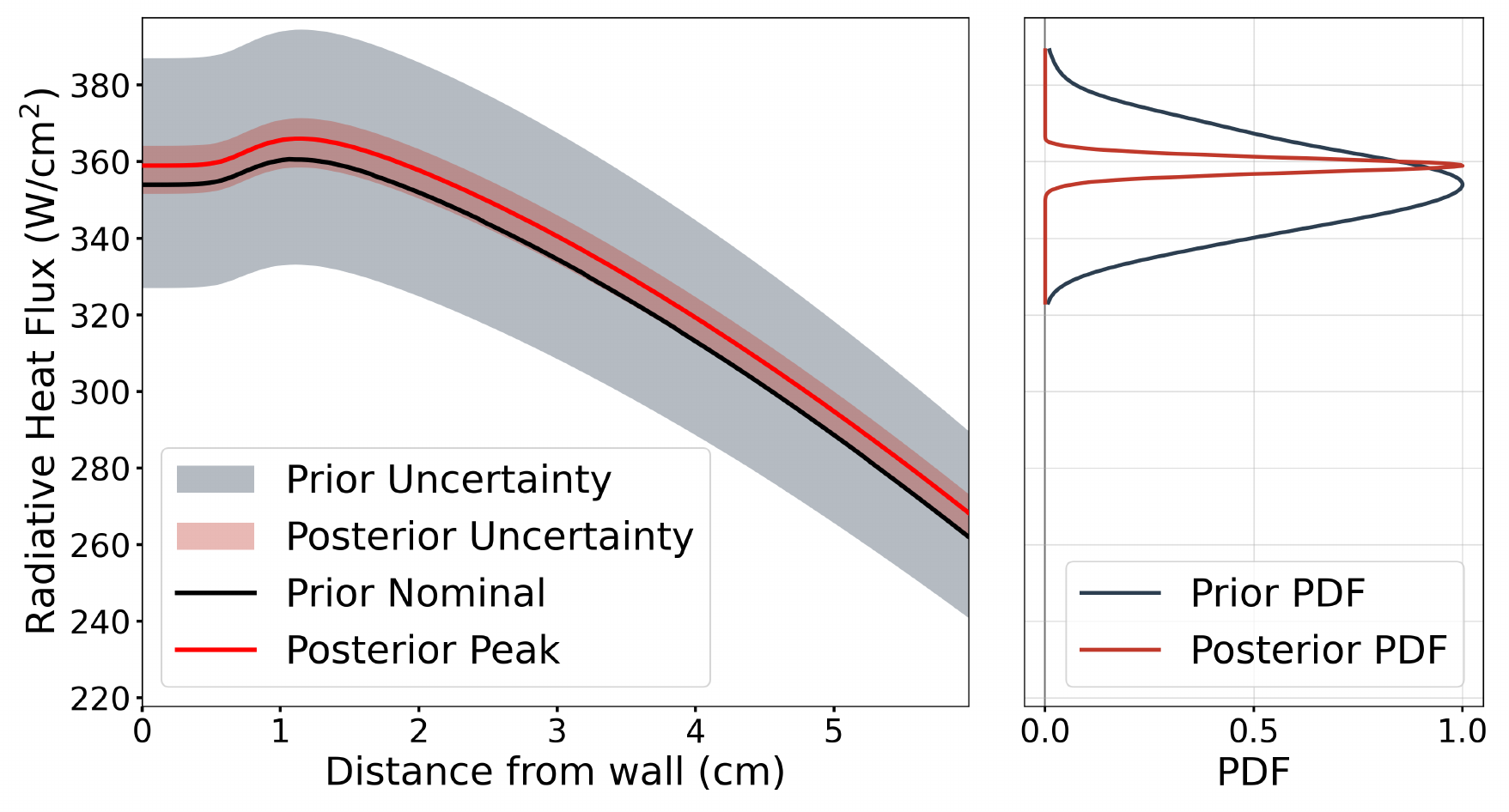}
    \caption{14 km/s.}
    \label{fig:fig15c}
\end{subfigure}
\caption{Non-Boltzmann radiative heat flux profiles along the stagnation line (left) and corresponding PDF slices at the stagnation point (right).}
\label{fig:fig15}
\end{figure}

Figure~\ref{fig:fig15} shows the predicted non-Boltzmann radiative heat flux distributions along the stagnation line, using the prior and posterior uncertainty ranges. In the left panel, the gray area denotes the expected variation of the radiative heat flux due to the prior parametric uncertainty in the Einstein and Stark broadening coefficients. Due to the prior uncertainty, the variation of the heat flux is significant, with the standard deviation (SD) of the heat flux reaching 10.4 W/cm${}^2$ at the entry speed of 14 km/s, as listed in Table \ref{tab:table5}. In contrast, through the Bayesian inference performed in the present study, the variation of the radiative heat flux is markedly reduced, as qualitatively shown in the left panels of Fig. \ref{fig:fig15} by the red area and quantitatively noted in Table \ref{tab:table5} in terms of SD. Across the different velocity conditions, the SD has been considerably reduced by a factor of approximately five through the present Bayesian inference. The reduction in predictive uncertainty for the radiative heat flux at the stagnation point is evident in the right panels of Fig. \ref{fig:fig15}. It is also worth noting that the radiative heat flux at the posterior peak (\emph{i.e.}, red line in the left panels) is higher than the prior nominal prediction, although the difference is limited to approximately 3\%.

\begin{table}[h]
\centering
\caption{Standard deviation (SD) of the radiative heat flux at the stagnation point.}
\label{tab:table5}
\begin{tabular}{c c c}
\hline\hline
\textbf{Velocity} 
& \textbf{SD with prior PDF (W/cm$^2$)} 
& \textbf{SD with posterior PDF (W/cm$^2$)} \\
\hline
10 km/s 
& $0.194$ 
& $0.0395$ \\
12 km/s 
& $3.490$ 
& $0.655$ \\
14 km/s 
& $10.40$ 
& $1.940$ \\
\hline\hline
\end{tabular}
\end{table}

\section{Conclusions}\label{Sec:Conclusions}
In the present study, the Einstein coefficients and Stark broadening coefficients of atomic nitrogen bound-bound transitions were inferred, and their parametric uncertainty was quantified. The inference was performed via Bayesian inversion of equilibrium spectral radiance measured in the NASA Ames Electric-Arc Shock Tube for shots T62-19 (10.32 km/s) and T62-21 (10.72 km/s). The analysis was restricted to the post-shock equilibrium region, where the Boltzmann assumption closes the species population degree of freedom; the residual uncertainty in the post-shock temperature and species number densities was then incorporated as a coupled nuisance parameter distribution. To make the inference tractable, a hybrid PCA/PCE surrogate model was constructed after Morris screening reduced each wavelength region to its most influential parameters. The likelihood was then formulated jointly over the two shots, with a tempering weight that down-weights the data contribution when the per-shot maximum a posteriori estimates disagree.

In total, eighteen parameters were inferred across the eight wavelength regions in which the prior model output envelops the EAST measurements: ten Einstein coefficients and eight Stark broadening coefficients. The resulting posterior uncertainties were significantly reduced compared with the prior literature bands. The joint posterior distributions were then forward propagated through the stagnation-line flow field around a 3 m radius sphere entering the Earth's atmosphere at 10, 12, and 14 km/s. Across all three entry conditions, the standard deviation of the predicted radiative heat flux was reduced by approximately a factor of five compared with the prior; at 14 km/s, it dropped from $10.4$ W/cm$^{2}$ to $1.94$ W/cm$^{2}$. The updated parameter set is therefore directly usable in radiation calculations for high-temperature atomic-nitrogen-rich shock layers, and it establishes a calibrated spectroscopic baseline for subsequent Bayesian inversion of kinetic processes such as heavy particle impact excitation. Extending the same framework to the diatomic nitrogen system, and to lower-shock-speed EAST conditions, is the natural next step.

\section*{Acknowledgments}
The work was partially supported by AFOSR Grant No. FA9550-25-1-0119 with Dr. Amanda Chou as Program Manager (to S.M.J.). The authors thank Dr. A. Munafò and Dr. M. Panesi (University of California, Irvine) for providing access to the \textsc{plato} and \textsc{hegel} software.

\bibliography{sample}

\end{document}

\maketitle
\renewcommand{\thefigure}{S\arabic{figure}}   
\renewcommand{\thetable}{S\arabic{table}}   
\renewcommand{\thepage}{S\arabic{page}} 

\begin{table}[htbp]
\caption{Grouped electronic energy levels and degeneracies.}
\label{tab:energy_levels}
\centering
\begin{tabular}{cccc}
\hline\hline
State for N & Energy, cm$^{-1}$ & Degeneracy & $n_p$ \\
\hline
1  & 0      & 4    & 2 \\
2  & 19228  & 10   & 2 \\
3  & 28839  & 6    & 2 \\
4  & 83336  & 12   & 3 \\
5  & 86193  & 6    & 3 \\
6  & 88132  & 12   & 2 \\
7  & 93582  & 2    & 3 \\
8  & 94838  & 20   & 3 \\
9  & 95510  & 12   & 3 \\
10 & 96751  & 4    & 3 \\
11 & 96833  & 10   & 3 \\
12 & 97794  & 6    & 3 \\
13 & 99664  & 10   & 3 \\
14 & 103694 & 12   & 4 \\
15 & 104196 & 6    & 4 \\
16 & 104628 & 6    & 3 \\
17 & 104720 & 28   & 3 \\
18 & 104849 & 14   & 3 \\
19 & 104852 & 12   & 3 \\
20 & 105007 & 20   & 3 \\
21 & 105134 & 10   & 3 \\
22 & 106478 & 2    & 4 \\
23 & 106823 & 20   & 4 \\
24 & 107014 & 12   & 4 \\
25 & 107225 & 10   & 4 \\
26 & 107446 & 4    & 4 \\
27 & 107615 & 6    & 4 \\
28 & 110021 & 18   & 5 \\
29 & 110315 & 90   & 4 \\
30 & 110486 & 126  & 4 \\
31 & 111363 & 54   & 5 \\
32 & 112851 & 90   & 5 \\
33 & 112929 & 288  & 5 \\
34 & 114298 & 648  & 6 \\
35 & 115107 & 882  & 7 \\
36 & 115631 & 1152 & 8 \\
37 & 115991 & 1458 & 9 \\
38 & 116248 & 1800 & 10 \\
\hline\hline
\end{tabular}
\end{table}